\documentclass[reprint,amsmath,amssymb,aps,showpacs]{revtex4-1}
\usepackage{graphicx}
\usepackage{dcolumn}
\usepackage{bm}
\usepackage{float}
\usepackage{epstopdf}
\usepackage{epsfig,color}
\usepackage{amsmath}
\usepackage{amsfonts}
\usepackage{amssymb}
\date{today}
\newcommand{\be}{\begin{equation}}
\newcommand{\ee}{\end{equation}}
\newcommand{\bea}{\begin{eqnarray}}  
\newcommand{\eea}{\end{eqnarray}}

\newcommand{\p}{\partial}
\newcommand{\s}{\sigma}

\newcommand{\up}{\uparrow}
\newcommand{\down}{\downarrow}
\newcommand{\la}{\langle}
\newcommand{\ra}{\rangle}
\newcommand{\rd}{\mbox{d}}
\newcommand{\ri}{\mbox{i}}

\newcommand{\eps}{\epsilon}
\newcommand{\nn}{\nonumber}

\newcommand{\bz}{\bar{z}}

\newcommand{\vxi}{\mbox{\boldmath $\xi$}}
\newcommand{\veta}{\mbox{\boldmath $\eta$}}
\newcommand{\vzeta}{\mbox{\boldmath $\zeta$}}

\newcommand{\htau}{\hat{\tau}}

\input{epsf}

\begin{document}

\title{Phase diagram 
of an interacting staggered
Su-Schrieffer-Heeger two-chain ladder close to a quantum critical point}
\author{A.A.Nersesyan$^{1,2,3}$}
\affiliation{\small {\sl $^1$The Abdus Salam International Centre for Theoretical 
Physics, 34151, Trieste, Italy}}
\affiliation{\small {\sl $^2$Tbilisi State University, Andronikashvili Institute of Physics, 0177, Tbilisi, Georgia,}}
\affiliation{\small{\sl $^3$Ilia State University, 0162, Tbilisi, Georgia,}}
\date{\today}

\begin{abstract}

We study the ground-state phase diagram of an
interacting staggered Su-Schrieffer-Heeger (SSH) ladder in the vicinity of the Gaussian 
quantum critical point.
The corresponding effective field theory which non-perturbatively treats correlations effects in the ladder
is a double-frequency sine-Gordon (DSG) model. It involves
two perturbations at the Gaussian fixed point: the deviation from criticality 
and  Umklapp scattering
processes. 
While massive phases with 
broken symmetries are identified by 
means of local order parameters, 
a topological distinction between thermodynamically
equivalent phases becomes only feasible when nonlocal fermionic fields,  parity and string order parameter, are included into consideration. 
We prove that
a noninteracting fermionic staggered SSH ladder
is exactly equivalent to a O(2)-symmetric model of two decoupled Kitaev-Majorana chains, or two 1D $p$-wave
superconductors. 
Close to the Gaussian fixed point the SSH ladder maps to 
an Ashkin-Teller like system when interactions are included.
Thus, the topological order in the SSH ladder is related to broken-symmetry phases of the associated quantum spin-chain degrees of freedom.
The obtained phase diagram includes a Tomonaga-Luttinger liquid state 
which, due to Umklapp processes, can become unstable
against either spontaneous dimerization or the onset of a charge-density wave (CDW).
In these gapped phases elementary bulk excitations are quantum kinks carrying the charge $Q_F = 1/2$.
For sufficiently strong, long-range interactions,
the phase diagram of the model exhibits a 
bifurcation of the Gaussian critical point into two outgoing
Ising criticalities. The latter
sandwich a mixed phase in which dimerization coexists with a site-diagonal CDW.
In this phase elementary bulk excitations  are 
represented by
two types of topological solitons carrying different fermionic charges, which continuously interpolate between 0 and 1.
This phase has also mixed topological properties with coexisting  parity and string order parameters.
\end{abstract}
\maketitle

\section{Introduction}

The Su-Schrieffer-Heeger (SSH) model \cite{ssh}
describes a one-dimensional Peierls insulator \cite{peierls} in terms of tight-binding fermions
whose hopping along the chain is characterized by alternating
nearest-neighbor amplitudes $t_{\pm} = t_0\pm \Delta /2$.  
It was introduced four decades ago almost simultaneously with a closely related field theory
of (1+1)-dimensional fermions coupled to a semiclassical scalar field with a soliton-like 
background -- the Jackiw-Rebby model \cite{JR}. In these seminal works 
it has been demonstrated that, for certain chains with a degenerate gapped ground state,
such as {\sl trans}-polyacetylene,
the excitations associated with topological defects and
edge states in finite samples
are characterized by fractionalization of charge 
\cite{JR, JS} and the related phenomenon of charge-spin separation \cite{ssh}.

Shortly after the 
SSH papers \cite{ssh}, a two-chain SSH ladder model was proposed
to explain soliton confinement in arrays of weakly coupled dimerized chains \cite{BM1,BM2}.
Nowadays low-dimensional objects like dimerized 
chains and ladders are being successfully manufactured and studied in cold atom systems on
optical lattices \cite{li1,rosch,meier,zhang}.
Current interest in two- and multi-chain SSH ladders and related systems, which include hybrid models that interpolate 
between the SSH and Kitaev's $p$-wave superconducting chain \cite{kitaev},
as well as Creutz-Hubbard and Kitaev ladders,
is strongly enhanced by the interest in
the studies of topological  phases of such 
objects \cite{bahri, nagaosa, sticlet, piga, cheon, li, vishvesh}.  
Boundary zero-mode states characterizing such phases are believed to play an important role because of their potential for quantum computation \cite{kitaev, alicea}. 

The SSH ladder 
is described by the Hamiltonian
\[
H = H_{0} + H_{\rm int},
\] 
where
\bea
H_{0} =  &-& \sum_{n\s} \left[t_0 + \frac{1}{2} \Delta_{\s} (-1)^n   \right]
\left(c^{\dagger}_{n\s} c_{n+1,\s} + h.c.  \right) \nn\\
&-& t_{\perp} \sum_{n\s} c^{\dagger}_{n\s} c_{n,-\s} 
\label{ham-latt}
\eea
is a one-particle Hamiltonian of 
spinless (e.g. fully spin polarized) noninteracting fermions 
which 
includes dimerization ($\Delta_{\s}$) and 
single-particle interchain hopping ($t_{\perp}$). Here $c^{\dagger}_{n\s}$, $c_{n\s}$ are 
second-quantized
operators of a fermion on the site $n$ of the chain labeled by $\s = \pm 1$. 
The average number of fermions
per single rung of the ladder is 1.
The model acquires features of a strongly correlated 1D Fermi system
when interaction between the fermions is included. If one accounts for
nearest-neighbor interaction only, $H_{\rm int}$ takes the form
\be
H_{\rm int} 
= U \sum_n \hat{n}_{n,\up} \hat{n}_{n,\down}
+ V \sum_{n\s} \hat{n}_{n,\s} \hat{n}_{n+1,\s}
\label{int}
\ee
where
$ \hat{n}_{n,\s} = c^{\dagger}_{n\s} c_{n\s}$ are fermionic occupation number operators, and
the coupling constants $U$ and $V$ parametrize the interchain and in-chain repulsion.
$H_{\rm int} $ may also incorporate longer-range interactions between the particles.
Indeed, in Fermi mixtures of ultracold atoms, the properties
of a lower-dimensional subsystem, such as a single chain or two-leg ladder, 
can be manipulated by tuning parameters in the
higher-dimensional species to which the lower-dimensional subsystem is coupled. This is a way
how long-range interaction in ladder Fermi systems on optical lattices can be generated
\cite{tsai, tsai1}.

\begin{figure}[hbbp]
\centering
\includegraphics[width=3.3in]{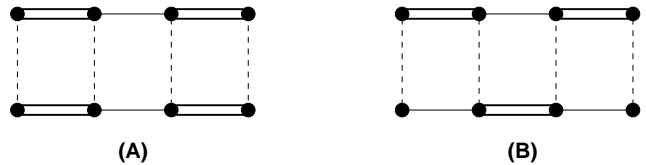}
\caption{\footnotesize Two dimerization patters of the ladder:
(A) columnar dimerization; (B) staggered dimerization.
The links with hopping amplitudes
$t_+$, $t_-$ and $t_{\perp}$
are depicted by the double, single and dashed lines, respectively.} 
\label{patterns}
\end{figure}

Fig.\ref{patterns} shows two paradigmatic  dimerization patterns:  (A) \emph{columnar} dimerization
with $\Delta_+ = \Delta_-$
and (B) \emph{staggered} dimerization with $\Delta_+ = -\Delta_-$.
In earlier theoretical \cite{BM1} and
experimental \cite{x-ray} studies is has been indicated that in quasi-1D systems 
of polyacetylene chains
already a weak interchain tunneling makes the staggered relative ordering of the chains
more stable. Apart from this, for purely theoretical reasons the B-type ladder appears to be
of particular interest.
In a columnar  ladder the role of the 
amplitude $t_{\perp}$ is
similar to that of the chemical potential in a usual two-band insulator. The
interchain hopping only controls the filling of the bands and thus can lead to insulator-metal 
(or commensurate-incommesurate \cite{c-ic1, c-ic2}) transitions,
without affecting the dispersion of the bands. At $t_{\perp} = 0$, the midgap states
realized in the bulk as a pair of solitons centered in the vicinity
of the same rung or, in the topological gapped phase, on the boundaries of the sample,
are doubly degenerate zero-energy modes, each carrying fractional charge $q_F = 1/2$.
These modes 
split into doublets due to interchain tunneling. As first shown by Baeriswyl and Maki 
\cite{BM1}, the two zero modes confine \cite{BM1,BM2}
to form a bound state
representing a single fermion 
with the charge $q_F = 1$. Thus, at any nonzero $t_{\perp}$ degenerate boundary modes and the associated fractional charge 
are no longer the property  of  a columnar ladder.

The situation is different in the staggered SSH ladder.  Here
$t_{\perp}$ couples to a non-conserved quantity 
which
makes the spectrum of the system 
essentially dependent of $t_{\perp}$ and similar to that in the Kitaev model
of a 1D p-wave superconductor \cite{kitaev}. 
However, there is a principal difference here: 
the global U(1) symmetry  of staggered dimerized ladder
leads to conservation of the total  particle number ${\cal N}$, whereas the Kitaev model has only a discrete symmetry $\mathbb{Z}_2$, "generated" by 
parity $P = (-1)^{\cal N}$.

What concerns bulk properties of the B-ladder, at any nonzero $\Delta$ it 
does not have a
metallic phase occupying a finite region in the parameter space. Instead, at $t_{\perp} = \pm 2t_0$ 
already a non-interacting staggered ladder displays two symmetric Gaussian critical points separating a topologically nontrivial massive phase ($|t_{\perp}| < 2t_0$)
from trivial phases ($|t_{\perp}| > 2t_0$). Due to chiral symmetry, the phase at
$|t_{\perp}| < 2t_0$ is 
topologically protected and characterized  by edge states with a fractional charge $q_F = 1/2$.
For non-interacting fermions, the charge of elementary bulk excitations is, of course, $Q_F = 1$.

In this work, we focus on the correlation effects in an
interacting staggered SSH ladder in the vicinity of the Gaussian criticality 
where the fermionic spectrum is gapless. Due to the $t_{\perp} \to -t_{\perp}$ symmetry,
we choose $t_{\perp}$ to be close to $2t_0$.
We derive
an effective low-energy field-theoretical model in which interaction is treated
nonperturbatively using Abelian bosonization. 
We show that close to the Gaussian criticality  interactions transform the ladder
to a strongly correlated 1D system and affect topological properties of massive phases. 
The phase diagram is rich and includes massive phases with an explicitly or spontaneously broken
discrete symmetries. Some of these phases are topologically nontrivial and some are not.
Exactly at $t_{\perp} = 2t_0$ the phase diagram displays a line of Gaussian critical points with continuously varying exponents
(Tomonaga-Luttinger liquid).
If interaction
is strong enough and/or is sufficiently long-ranged, Umklapp processes
make the Tomonaga-Luttinger critical state unstable against \emph{spontaneous} breakdown of either
link or site parity, leading to the onset of dimerization long-range order or the onset of a charge-density wave (CDW),
respectively. In these gapped phases elementary bulk excitations are not fermions but quantum kinks carrying the charge $Q_F = 1/2$.

To have a reliable tool to distinguish between topologically 
distinct
phases of the system
one needs a \emph{local} representation of \emph{nonlocal} fermionic fields: parity and string order parameter. The bosonization approach supplies
these nonlocal fields with a local representation.
To put this correspondence on firm ground we need to establish contact
between the ladder model and hidden Ising degrees of freedom in a symmetry preserving way.
We first demonstrate
that, in any range of its parameters, a non-interacting 
staggered SSH ladder can be \emph{exactly} mapped onto an O(2)-symmetric model of
two decoupled Kitaev-Majorana (KM) chains, or equivalently two copies of an XY spin-1/2 chain in a transverse
magnetic field, or two decoupled 1D $p$-wave superconductors (1DPS). 
Bearing in mind that our ladder represents a system of two SSH chains coupled
by interchain tunneling ($t_{\perp} \neq 0$), the possibility of such factorization appears as a remarkable property of the model.
The continuous O(2) symmetry shows up as  the invariance of the two KM chains under rotations of
the two-component Majorana vector field.

In the vicinity of the critical point,
the non-interacting ladder is equivalent to a pair of identical, weakly off-critical, decoupled quantum Ising chains. 
The fermionic nonlocal operators (parity and string order parameter)
are then identified as products of two order or disorder Ising parameters. In this way,
topological order in the ladder system is related to broken-symmetry phases of the associated
quantum spin-chain  degrees of freedom.
Switching on interaction between the original fermions on the ladder transforms the two decoupled Ising chains
to an quantum Ashkin-Teller model \cite{yama, delfino}. 
{The proof of this equivalence is one of the main results of the paper.}

Here a remark is in order.
When the staggered SSH ladder is considered in the vicinity of the critical point $t_{\perp}=2t_0$,
the existence of the aforementioned equivalence does not come as a revelation.
The theory of a massive Dirac fermion with a marginal interaction, the so-called massive Thirring model, has long been known to be equivalent to the quantum Ashkin-Teller system of two, marginally coupled
quantum Ising chains \cite{ZI}. On the other hand, the two theories are related to the quantum 
sine-Gordon model \cite{coleman}.
Main bosonization formulas exploring this triad of equivalence have been derived, including those
which concern nonlocal fermionic fields \cite{{dm},{dsg}}. 
Obviously, the significance
of the mapping of the fermionic staggered SSH ladder onto a O(2) theory of KM chains
follows from 
the fact that all the models involved are defined on a lattice and the mapping is exact.

We show that universal low-energy properties of the model are formed due to the interplay of two relevant perturbations: the deviation from the U(1) criticality and Umklapp processes
generates by interactions. Upon bosonization, such interplay is adequately described by
the quantum double-frequency sine-Gordon (DSG) model \cite{dm,dsg}.
This theory predicts realization of a typical Ashkin-Teller scenario: at small
deviations from the Gaussian criticality, 
the phase diagram of the model exhibits a 
bifurcation of the Gaussian critical point (central charge $c=1$)  into two outgoing
$Z_2$, or Ising criticalities (each with central charge $c=1/2$). The two
Ising critical lines sandwich a mixed phase in which dimerization coexists with a site-diagonal CDW.
In this phase, due to the CDW ordering, the charge conjugation symmetry  is spontaneously broken and, as a consequence, the fermionic number $Q_F$ is not quantized 
in units 1/2 (see e.g. Ref.[\onlinecite{ns}]). Elementary bulk excitations in the mixed phase are 
represented by
two types of topological solitons carrying different fermionic charges, which continuously interpolate between 0 and 1.
This phase has also mixed topological properties with continuously varying parity and string order parameters.

The paper is organized as follows.
In Sec.II we overview the spectral properties of a noninteracting staggered SSH ladder. In particular
we discuss
the evolution of the fermionic spectrum  on approaching
the critical point $t_{\perp} = 2t_0$.
On decreasing the parameter $\gamma = 1 - {t_{\perp}}/{2t_0} ~(|\gamma| \ll 1)$, at small
$\delta = \Delta/2t_0$,
we observe a smooth crossover between an incommensurate massive phase, $\gamma > \delta^2$,
and a commensurate massive phase, $\gamma < \delta^2$ (the commensurate
phase extends to the region $\gamma < 0$). In the latter case, the elementary excitation
represents a Dirac-like fermion with a mass $m \sim \gamma$. 
At $m=0$ one has
a continuum theory of a massless fermion with a \emph{single} Fermi point at $k=0$.
In Sec.III we incorporate interactions between the fermions into an effective continuum
model and then bosonize it.
 As a result, we arrive at
the DSG model where the original Dirac mass term
and Umklapp processes are the key pertubations to the Gaussian scalar field theory. Here we also derive
the bozonized form of all local physical fields.

In Sec.IV an exact equivalence between the staggered SSH ladder and a pair of Kitaev chains 
is established. 
Close to the U(1) criticality
($|\gamma| \ll 1$) 
interactions transform this system to a quantum Ashkin-Teller model, which makes
it possible to employ the previously developed formalism that derives the low-energy projections of all physical fields of the DSG model in terms of the constituent spin degrees of freedom \cite{dsg}.
This equivalence proves instrumental to derive bosonized expressions for nonlocal fermionic operators, parity and string order parameter. In Sec.V we provide a local, field-theoretical representation of the
parity an string-order operators, together with their representation in terms of the Ising variables.
In Sec.VI we discuss in much detail the ground state phase diagram of the staggered
ladder paying attention to quantum critical lines separating massive phases, topological properties of the latter and the fermionic numbers carried by elementary excitations.
The paper has two appendices where some details of Abelian bosonization and basic facts about the Kitaev-Majorana model are compiled.

\section{The spectrum of noninteracting staggered SSH ladder}

We start our discussion by overviewing main properties of a noninteracting staggered SSH ladder.
While the  columnar ladder is symmetric 
under the interchange $P_{12}$ of the two chains, 
the
Hamiltonian of the staggered ladder
has a glide reflection symmetry \cite{zhang}
which is a direct product of  a translation by one lattice spacing (${\cal T}_a$)
and reflection ($P_{12}$). 
The spectrum of the staggered ladder 
remains fully defined within the original Brillouin zone $|k|<\pi$.

Passing at each rung to bonding ($b$) and antibonding ($a$) states
\be
a_k = \frac{1}{\sqrt{2N}} \sum_{n,\s} \s c_{n\s} e^{-ikn}, ~
b_k = \frac{1}{\sqrt{2N}} \sum_{n,\s} c_{n\s} e^{-ikn}
\ee
we represent the Hamiltonian (\ref{ham-latt}) at $\Delta_+ = - \Delta_- \equiv \Delta$
as follows:
\bea
&& H_0 = \sum_{|k|<\pi} \psi^{\dagger}_k \hat{h}(k) \psi_k,~~~~
\psi_k = \left(
\begin{array}{clcr}
a_k \\
b_{k+\pi}
\end{array}
\right),\nn\\
&&~~~~\hat{h}(k) = \left( \eps_k + t_{\perp} \right)\htau_3 - \Delta_k \htau_2
\label{h-B}
\eea
where the Pauli matrices $\htau^a ~(a=1,2,3)$ act in the two-dimensional Dirac-Nambu space.
Everywhere below we will assume that 
$t_0 > 0$,  $0 < \Delta \leq 2t_0$, while the ratio $t_{\perp}/2t_0$
is arbitrary.
The model (\ref{h-B}) has a chiral symmetry,
$\psi_k \to \htau_1 \psi_k$, ~$\htau_1 \hat{h}(k) \htau_1 = -  \hat{h}(k)$,
implying that the spectrum of the Hamiltonian consists of ($E,-E$) pairs 
\bea
E^{(\pm)} (k) = \pm E(k), ~~~E(k) = \sqrt{(\eps_k + t_{\perp})^2 + \Delta^2 _k}
~~~\label{B-spec-fin}
\eea
and possibly contains
zero-energy modes.

\begin{figure}[hbbp]
\centering
\includegraphics[width=1.4in]{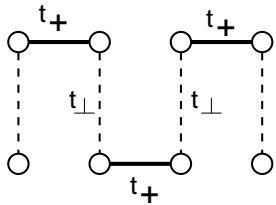}
\caption{\footnotesize Staggered ladder
at $t_- = t_0 - \frac{1}{2}\Delta = 0$, $t_+ = 2t_0$.}
\label{special-Plad}
\end{figure}

The Hamiltonian $H_0$ conserves the total charge ${\cal N} = \sum_k \psi^{\dagger}_k \psi_k$
but because of interband transitions caused by 
dimerization does not conserve
the "chiral charge" ${\cal N}_3 = \sum_{k} \psi^{\dagger}_k \hat{\tau}_3 \psi_k$. This is why, similar to
the chemical potential in a BCS superconductor, $t_{\perp}$ appears in (\ref{B-spec-fin})
inside the square root.
Therefore there is no room for quantum commensurate-incommensurate
transitions \cite{c-ic1,c-ic2} in this case.
In fact, the B-ladder does not possess a metallic phase 
extending over a finite range of $t_{\perp}$. Instead at any fixed $\Delta \neq 0$ the 
spectrum (\ref{B-spec-fin}) remains gapped except for
two isolated critical points occurring at 
\bea
k=0: ~~~ t_{\perp} = 2t_0 ~~~~~\textrm{(upper critical point)}\label{upper}\\
k=\pi:~~~ t_{\perp} = - 2t_0  ~~~\textrm{(lower critical point})\label{lower}
\eea
These points separate
massive phases that occupy the regions $|t_{\perp}| \neq 2t_0$.
The criticalities belong to the universality class of a free massless fermion:
Gaussian U(1) criticality with central charge $c=1$. 
The existence
of this criticality is immediately understood
in the special case 
\be
\Delta = 2t_0, ~t_{+} = t_0 + \frac{\Delta}{2} = 2t_0, ~t_{-} = t_0 - \frac{\Delta}{2} = 0
\label{special-choice}
\ee
(or equivalently $\Delta = - 2t_0$, $t_+ = 0,~t_- = 2t_0$) displayed in
Fig.\ref{special-Plad}.
The staggered ladder transforms to a single chain
with alternating 
hopping amplitudes $2t_0$ and $t_{\perp}$.  
Generically, the spectrum of such chain is massive;  however, at
$t_{\perp} = 2t_0$ translational invariance is restored, and
the resulting snake-looking uniform chain with a 1/2-filled tight-binding
band has a gapless spectrum. A similar situation is known to exist in the theory
of explicitly dimerized  spin-1/2 Heisenberg ladders \cite{snake1,snake2,wn}.
Notice that 
the conditions (\ref{upper}), (\ref{lower})
are less restrictive than those
corresponding to the translationally invariant snake-ladder of Fig.\ref{special-Plad}. 
In the latter case the critical points are determined by two conditions imposed on both
$t_{\perp}$ and $\Delta$, whereas (\ref{upper}) or (\ref{lower}) represents one condition
imposed on $t_{\perp}$ only.
Therefore, on the phase plane ($\Delta, t_{\perp}$) there exists critical \emph{lines} 
$t_{\perp} = \pm 2t_0$ 
along which $\Delta$ may be varied.

The spectrum (\ref{B-spec-fin}) of the noninteracting B-type SSH ladder
coincides with that of a
one-dimensional spinless superconductor with a $p$-wave pairing (1DPS) -- 
the Kitaev model \cite{kitaev, alicea}
(see Appendix \ref{kitaev}).
This similarity between the
staggered SSH ladder and the Kitaev model has been mentioned 
in the literature and topological properties of the two models were compared \cite{vishvesh}.
In both models the critical points $t_{\perp} = \pm 2t_0$ separate topologically nontrivial
massive phases ($|t_{\perp}| < 2t_0$) from trivial massive phases ($|t_{\perp}| > 2t_0$).
By the bulk-boundary correspondence \cite{HK}, in both cases the topological phase shows up in the appearance of
boundary zero-energy midgap states.
However, there is an important difference. 
In the Kitaev model 
the global symmetry  is $\mathbb{Z}_2$.  Therefore 
the boundary states localized at the edges of a single Kitaev chain are Majorana zero modes \cite{kitaev,alicea}.
These 
modes constitute 
a highly nonlocal realization of a 
$\mathbb{Z}_2$-degenerate
many-fermion ground state of a 1D
$p$-wave superconductor, characterized by even and odd parity of the particle number. On the other hand,
the continuous U(1) symmetry of the staggered SSH ladder
leads to conservation of the total particle number. 
Therefore in the topologically nontrivial phase of the noninteracting staggered ladder 
the two degenerate boundary Majorana modes combine to produce a zero-energy state of a 
\emph{complex} fermion 
carrying a fractional fermion number\cite{ssh} $q_F = 1/2$. The situation in the interacting ladder
will be discussed in Sec.\ref{phase-dia}.

The aforementioned differences make a direct
mapping of the SSH ladder onto a single $p$-wave superconducting chain illegitimate.
In  Sec.\ref{equiv} we demonstrate that, in the absence of interaction, the staggered 
ladder with two SSH chains coupled by interchain tunneling is
exactly equivalent to two \emph{decoupled} Kitaev chains. 
Apparently, the symmetry of such system is $\mathbb{Z}_2 \times \mathbb{Z}_2$. However,
passing to a Majorana representation of the two-chain Kitaev model reveals its
invariance under global
O(2) rotations of the two-component Majorana vector field, which 
correctly reproduces the U(1) symmetry of the original SSH ladder model.

Let us now
derive the effective fermionic 
Hamiltonian which captures the  low-energy properties of the noninteracting model
in the vicinity of the upper critical point (\ref{upper}), $t_{\perp} = 2t_0$. The lower critical point
(\ref{lower}),
$t_{\perp} = - 2t_0$ can be accessed using the symmetry
$
E(k, -t_{\perp}) = E (\pi - k, t_{\perp}).
$
Introducing two smooth fields slowly varying over the lattice constant 
$a_0$ ~($\Lambda \lesssim 1/a_0$),
\bea
&&\psi_a (x) = \frac{1}{\sqrt{L}} \sum_{|k| < \Lambda} e^{ikx} a_k, \nn\\ 
&&\psi_b (x) = \frac{1}{\sqrt{L}} \sum_{|k| < \Lambda} e^{ikx} b_{k+\pi}
\label{psi-ab-cont}
\eea
we can write the Hamiltonian density as
\bea
&&{\cal H}^{(0)} (x) =
\Psi^{\dagger} (x) \left[ \ri v_0 \p_x \hat{\tau}_2 - \left(m  + \frac{1}{2m^*}\p^2 _x\right)
\hat{\tau}_3
\right]
\Psi(x) + \cdots , \nn\\
&& ~~~~~~~~~~~~~~~~~~\Psi(x) = \left(
\begin{array}{clcr}
\psi_a (x)\\
\psi_b (x)
\end{array}
\right)
\label{H-cont-x}
\eea 
Here the dots stand for higher-order gradient terms.
The parameters in (\ref{H-cont-x} ) are
\bea
&&m = 2t_0 - t_{\perp} \equiv 2t_0 \gamma, ~~v_0 = \Delta a_0 = v_F \delta, \nn\\
&& ~~~~~~~~~~~~~~~~{1}/{2 m^*} = t_0 a^2 _0
\label{parameters}
\eea
$v_F = 2t_0 a_0$ being the Fermi velocity of the fermions on a single undimerized chain.  
(\ref{H-cont-x}) is the Hamiltonian density of free (1+1)-dimensional fermions which, apart from the
Dirac mass $m$, includes a non-relativistic correction $(m^{*})^{-1}\p^2 _x$.
The spectrum of $\hat{\cal H}^{(0)}$ represents a small-$k$ expansion
\bea
E^2 (k) &=& k^2 v^2 _0 +  \left(m  - \frac{k^2}{2m^*}\right)^2 \nn\\
&=& m^2 + k^2 v^2 + \frac{k^4}{4m^{*2}} + \cdots \label{small-k-spec}
\eea
where
\bea
v^2  = v^2 _0 - \frac{m}{m^*}
= v^2 _F (\delta^2 - \gamma)
\label{v2}
\eea
The $k^4$-term in (\ref{small-k-spec}) 
plays an important role in the formation of
incommensurate spatial correlations of the physical quantities  not too close to the
critical point (the subcritical regime $\delta^2 < \gamma  \ll 1$).
\begin{figure}[hbbp]
\centering
\includegraphics[width=3.2in]{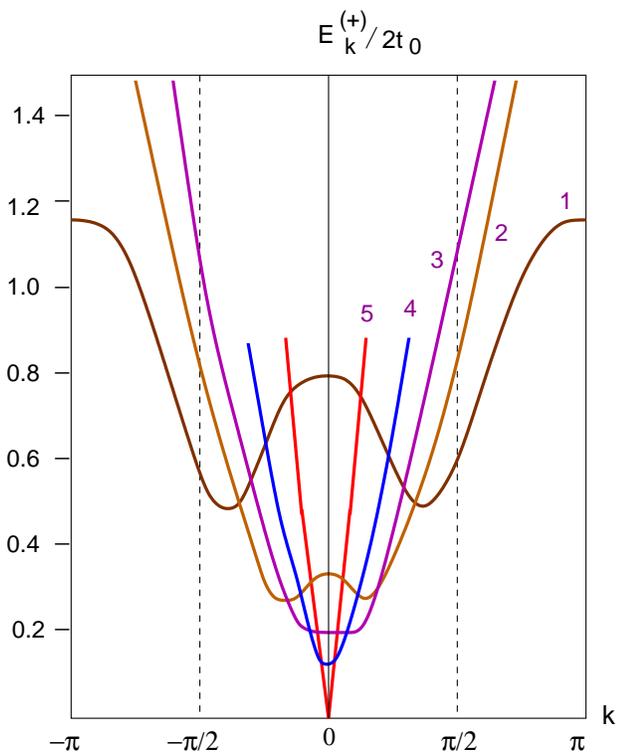}
\caption{\footnotesize The spectrum $E^{(+)}_k$ of the staggered ladder.
Notations: $\tau = t_{\perp}/2t_0$, $\delta = \Delta/2t_0 $. 
Chosen: $\delta^2 = 0.2$. Cases: (1) $\tau=0.1$, \\(2) $\tau = 0.7$,
(3) $\tau = 0.8$, (4) $\tau = 0.9$, (5) $\tau = 1$.}
\label{B-spec}
\end{figure}

Let us now discuss the consequences following from the specific form of the
one-particle spectrum (\ref{small-k-spec}) in the vicinity of the critical point, $|\gamma| \ll 1$
(see Fig.\ref{B-spec}).
We  assume that $0<\delta \ll 1$.
Depending on the sign of $v^2$ in
the 
expansion (\ref{small-k-spec}) 
there are two  regimes within the gapped phase $0<\gamma \ll 1$: 
$(\ri)$  massive \emph{incommensurate} regime: 
{$v^2 < 0$, $\delta^2 < \gamma \ll 1$};
$(\ri \ri)$ massive \emph{commensurate} regime: 
{$v^2 > 0$, $\gamma < \delta^2 \ll 1$}.
In the regime $(\ri)$, ${E} (k)$ 
has two symmetric minima at $k = \pm k_0$, where $k_0  = a^{-1}_0\sqrt{2 (\gamma - \delta^2)}$.
These minima, seen in the curves 1 and 2 of Fig.\ref{B-spec}, evolve from the two original minima at 
the Fermi momenta $\pm k_F = \pm \pi/2a_0$  
where a spectral gap opens up 
in the limit of two decoupled SSH chains ($t_{\perp}=0$).
On decreasing $\gamma$ the momentum $k_0$ decreases and vanishes
at the point $\gamma =\delta^2$ where $E(k) \simeq 2t_0 \gamma \left[ 1 + (ka_0)^4 / 8 \gamma^2 \right]$
(curve 3 in Fig.\ref{B-spec}).
Further decreasing $\gamma$ makes $v^2$ 
positive, 
and the model crosses over to the massive region $(\ri\ri)$
in which 
the dispersion curve has only one minimum at $k=0$ (curve 4 in Fig.\ref{B-spec}).
The $k^4$-term in the expansion (\ref{small-k-spec}) can be neglected 
under the condition that
$
|k|a_0 \ll \sqrt{\delta^2 - \gamma}.
$
Then one arrives at the spectrum of
a massive Dirac fermion $E(k) \simeq \sqrt{k^2 {v}^2 + {m}^2}$.
Exactly at the critical point $t_{\perp} = 2t_0 ~(\gamma = 0)$ the fermion becomes massless:
$
E(k) = v |k|.
$

If $\gamma$ is negative ($t_{\perp} > 2t_0, ~m<0$), ${v}^2$ 
remains positive,
and the dispersion curve always has a {single} minimum at $k=0$.
So at small but negative $\gamma$ 
one has the spectrum of a massive Dirac 
fermion.

The appearance of two spectral minima of $E(k)$ at $k = \pm k_0$ indicates that
in the region ($\ri$) spatial correlations of local physical fields must 
exhibit  incommensurate modulations with the period $2\pi / 2k_0$.
On the other hand, since the spectrum is gapped, these correlations should fall off exponentially
at distances larger than the correlation length $\xi_0$. The study of this question, which
will include computation of
spatial density-density correlation functions in both incommensurate and commensurate massive regimes, will be postponed until
a separate publication \cite{tn}. Here we would only like to stress that,
in the staggered SSH ladder,
crossing the point $\gamma = \delta^2 ~(k_0 = 0)$,
does not have a character of a phase transition.  It rather signifies a smooth crossover
between the massive  regimes.

\section{Including interactions}

Now we turn to 
interaction between the fermions as described by $H_{\rm int}$
in 
Eq.(\ref{int}). Naturally, correlation effects are expected to be most strongly pronounced
in the  vicinity of the critical points ($|t_{\perp}| \sim  2t_0$). 
To derive a continuum representation of $H_{\rm int}$, 
we will ignore the $k^4$-correction to the single-particle spectrum (\ref{small-k-spec}) 
and proceed from the "relativistic" model of a massive Dirac fermion,  Eq.(\ref{H-cont-x})
with the "nonrelativistic mass" $m^*$ sent to infinity.
The  SU(2) "spin" symmetry of the Hubbard on-site interaction implies that $H_{U}$
is invariant under rotations in the chain 
space. Therefore
\be
H_{ U} = 
g_U \int \rd x~\psi^{\dagger}_a (x) \psi_a (x) \psi^{\dagger}_b (x) \psi_b (x) 
\label{U-cont-limit}
\ee
where $g_U = Ua_0$ is the coupling constant. Furthermore,
using the correspondence
\[
c^{\dagger}_{n\s} c_{n\s} ~\to~
\frac{a_0}{2} \Big[
\Psi^{\dagger}(x) \Psi(x) + \s (-1)^n \Psi^{\dagger}(x) \htau_1\Psi(x)\Big]
\]
we find that
\bea
H_{V} 
&& = \frac{g_{V}}{2} \int \rd x~
\Big[ \left( \Psi^{\dagger}(x) \Psi(x)  \right)^2 \nn\\
&&-
\left( \Psi^{\dagger}(x) \htau_1\Psi(x)  \right) 
\left(  \Psi^{\dagger}(x+a_0) \htau_1\Psi(x+a_0) \right)\Big]
~~~\label{V-inter-cont}
\eea
where
$
g_{V} = Va_0
$
is another coupling constant. 
It is convenient to make a chiral rotation of the spinor $\Psi$:
\bea
\Psi (x) =  e^{- i \pi \hat{\tau}_1 /4} \chi (x), 
~~~~\chi(x) = \left(
\begin{array}{clcr}
R(x)\\
L(x)
\end{array}
\right)
\label{chiral-rot}
\eea
Under this rotation $\htau_2 \to - \htau_3$, $\htau_3 \to \htau_2$ and
the effective Dirac Hamiltonian of free massive fermions becomes
\bea
{\cal H}_{0} (x) 
= \chi^{\dagger} (x) \left[ - \ri v_0 \p_x \hat{\tau}_3 - m  
\hat{\tau}_2
\right]
\chi(x) \label{dirac-canonical}
\eea
To find the continuum form of the interaction in the 
(RL)-basis of
single-particles states we use the relations

\bea
:\psi^{\dagger}_{a(b)}  \psi_{a(b)} :~=~ \frac{1}{2} \left( J_R + J_L  \right) \mp 
\frac{\ri}{2}
\left( R^{\dagger} L - L^{\dagger} R\right); \nn
\eea
\bea
:\psi^{\dagger}_a  \psi_a : :\psi^{\dagger}_b  \psi_b :~
=~~\frac{1}{4} \left( J^2 _R + J^2 _L \right)
+ J_R J_L \nn\\
+  \frac{1}{4} \left[ \left( R^{\dagger} L \right)_x  \left(R^{\dagger} L\right)_{x+a_0} + h.c. \right]
\eea
where
$
J_R=:R^{\dagger} R:
$
and
$ J_L =:L^{\dagger}  L:$
are normal ordered densities of the right and left fermions, i.e. the U(1) chiral currents
(see Appendix \ref{bos}).
Taking into account the fact that $H_{V}$ maintains its structure with $\Psi$
replaced by $\chi$, we arrive at the following expression for the interaction density,
which is parametrized by
two coupling constants
\begin{widetext}
\bea
{\cal H}_{\rm int} = \frac{1}{2} g_+ (J_R ^2 + J_L ^2)
+ 2g_+  J_R J_L + \frac{1}{2} g_- 
\left[ \left( R^{\dagger} L \right)_x  \left(R^{\dagger} L\right)_{x+a_0} + h.c.   \right], ~~~
g_{\pm} = (g_U \pm 2 g_V)/2
\label{tot-int-cont}
\eea
\end{widetext}
The first term in the r.h.s. of (\ref{tot-int-cont}) renormalizes
the group velocity of the collective excitations, the second term is a marginal 
forward-scattering part of the
interaction, and the last term 
describes 
Umklapp processes whose correct treatment in a continuum field theory of spinless fermions requires point 
splitting \cite{haldane-xxz}.

Now we apply the bosonization method 
to the continuum fermionic model ${\cal H}_{\rm eff} (x) = {\cal H}_{0} (x) + {\cal H}_{\rm int} (x)$,
where ${\cal H}_{0}$ and ${\cal H}_{\rm int}$ are given by Eqs.(\ref{dirac-canonical})
and (\ref{tot-int-cont}), respectively. The details of this derivation can be found in
the Appendix \ref{bos} where the main steps  of Abelian bosonization are briefly outlined.
The bosonic counterpart of ${\cal H}_{\rm eff} (x)$
represents a double-frequency sine-Gordon (DSG) 
model \cite{dm,dsg}: 
\bea
{\cal H}_{\rm DSG} &=&  \frac{u}{2} \left[ \pi^2 (x) + \left(\p_x \phi(x)   \right)^2   \right]
+ \frac{m}{\pi \alpha} \cos \sqrt{4\pi K} \phi \nn\\
&-& \frac{\tilde{g}}{2(\pi\alpha)^2} \cos \sqrt{16 \pi K} \phi
\label{DSG2}
\eea
where 
\be
K = 1 - \frac{g_+}{\pi u} + O(g^2)   
\label{K-const}
\ee
The first term in the r.h.s. of (\ref{DSG2}) describes a conformally invariant Gaussian model
with central charge $c=1$. $\phi(x)$ and  $\pi(x)$ are the massless scalar field
and its conjugate momentum, respectively, $u$ being the renormalized velocity of collective
excitations.
We remind that the "Dirac mass" $m \simeq 2t_0 \gamma$ measures the deviation from the critical point 
($\gamma = 0$).
The mass and Umklapp terms represent two perturbations with
Gaussian scaling dimensions 
$d_{1} = K$ and $d_{2} = 4K$, respectively.

At a weak short-range repulsive interaction the Luttinger-liquid parameter $K$ is only slightly
less than 1, and the Umklapp term  in (\ref{DSG2}) is strongly irrelevant ($d_2 > 2$) 
at the Gaussian fixed point.
Since the DSG model is non-integrable \cite{dm}, the exact dependence of the Luttinger-liquid parameter $K$ on
the coupling constants is not known. 
To remedy  this shortcoming
we  can imagine that our SSH ladder model 
incorporates longer-range interactions which push this parameter to smaller values \cite{giam}, including 
$K = 1/2$. 
Below this value
both the mass and Umklapp terms become relevant and the interplay of the two perturbations
can lead to new infrared physics.

The unrenormalized Umklapp coupling constant $\tilde{g}$ is proportional to $g_-$ and 
changes its sign
at $g_U = 2g_V$. Even though in the effective infrared theory the precise
dependence of $\tilde{g}$  on the bare interaction constants is not universal, the exists
a line in the $UV$ plane where $\tilde{g}$ changes its sign \cite{fradkin}. This fact is crucial
for the physical consequences about the phase diagram of the model.

The DSG model  (\ref{DSG2}) must be supplemented by the list of bosonized strongly fluctuating
\emph{local} fields.

\noindent
1) The total on-rung
density fluctuation is defined as
\bea
&&:\rho(n):  = \sum_{\s} :c^{\dagger}_{n\s} c_{n\s}: 
\to
a_0 J(x), \nn\\
&& J(x)= :\Psi^{\dagger} (x) \Psi(x): = :\chi^{\dagger}(x) \chi(x):\nn\\
&&~~~~~~\to
\sqrt{{K}/{\pi}}~ \p_x \phi(x)
\label{totden}
\eea
and the fermion number is defined as 
\bea
&&Q = \int_{-\infty}^{\infty} \rd x~J(x) = 
\sqrt{{K}/{\pi}}~ \Delta \phi, \label{Q-top}\\
&&~~~~~~~~~~\Delta \phi = \phi(\infty) - \phi(-\infty)
\nn
\eea
2) The total longitudinal and transverse bond-densities, both measured from their ground state average values
at the critical point $\gamma = 0$, are
\bea
:{\cal D}_{n,n+1}: &=& (-1)^n \sum_{\s}\left( :c^{\dagger}_{n\s} c_{n+1, \s}:~ + h.c.  \right)\nn\\
 &\to& a_0 {\cal B}_{\parallel} (x), \nn\\
:{\cal B}_{\perp}(n): &=& \sum_{\s} :c^{\dagger}_{n\s} c_{n,-\s}:
~\to~
a_0 {\cal B}_{\perp}(x)\nn\\
{\cal B}_{\parallel} (x) &\sim& -   {\cal B}_{\perp}(x) \sim :\Psi^{\dagger} (x) \htau_3 \Psi(x):
= \chi^{\dagger} (x) \htau_2 \chi(x) \nn\\
&\to& - (\pi\alpha)^{-1}:\cos \sqrt{4\pi K} \phi(x):
\label{Dbos}
\eea

\noindent
3) The staggered part of the  site diagonal relative density (charge density wave, CDW) transforms to
\bea
\s(n) &=& (-1)^n \sum_{\s} \s :c^{\dagger}_{n\s} c_{n\s}: ~\to~
a_0 \rho_{\rm CDW} (x), \nn\\
~~\rho_{\rm CDW} (x) &=& :\Psi^{\dagger} (x) \htau_1 \Psi(x):
= \chi^{\dagger} (x) \htau_1 \chi(x) \nn\\
 &\to& - (\pi\alpha)^{-1} :\sin \sqrt{4\pi K} \phi (x):
\label{relatden}
\eea

\noindent
In Sec.\ref{nonlocalOP} we show that the above list must be complemented by two more
operators,
$\cos \sqrt{\pi K}\phi(x)$
and $\sin \sqrt{\pi K}\phi(x)$, which represent 
nonlocal fermionic fields: parity and string order parameter. The latter play a crucial role
in identifying topologically non-trivial massive phases of the interacting system.

\section{Equivalence between staggered SSH ladder
and a pair of Kitaev chains} \label{equiv}

\begin{figure}[hbbp]
\centering
\includegraphics[width=2.4in]{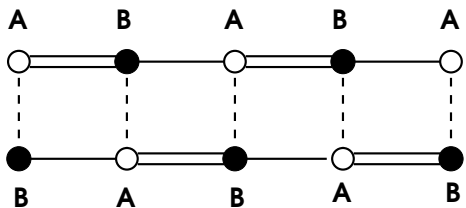}
\caption{\footnotesize Two-sublattice representation of the staggered SSH ladder.} 
\label{2sublat}
\end{figure}

In this section we 
establish
an exact equivalence  
between the staggered SSH ladder
and a pair of 
 Kitaev chains. 
Let us divide the lattice of the staggered ladder
into A and  B sublattices  as shown in Fig.\ref{2sublat}.
Associated with these sublattices are
two fermionic operators,  $\alpha_n$ and $\beta_n$ which are   
related to the original operators $c_{n\s}$ as follows:
\[
c_{n\s} = (-1)^n \left[ e^{i\pi/4}  \Pi_{\s} (n) \alpha_n -  e^{-i\pi/4} \Pi_{-\s} (n) \beta_n
\right]
\]
Here $\Pi_{\pm} (n) = [1 \pm  (-1)^n]/{2}$ are projectors on the even and odd sites, respectively.
We then rewrite the Hamiltonian of the noninteracting staggered SSH ladder
in a translationally invariant form
\bea
H_0  &=&  \ri \sum_n \left( t_+ \alpha^{\dagger}_n \beta_{n+1} + 
t_-  \alpha^{\dagger}_n \beta_{n-1}
- t_{\perp} \alpha^{\dagger}_n \beta_n \right) \nn\\
&+& h.c.\label{ham-rep-mod}
\eea
The spectrum of this Hamiltonian, $ E^{(\pm)}(k)$, coincides with the expression
(\ref{B-spec-fin}), as it should.
Splitting the operators $\alpha_n$ and $\beta_n$ 
into pairs of Majorana (real) fermions, $\left( \eta^a _n, \zeta^a _n  \right)$
\bea
&& ~~~\alpha^{\dagger}_n = \frac{\eta^{(1)}_n + \ri \eta^{(2)}_n}{2}, ~~
\beta^{\dagger}_n = \frac{\zeta^{(1)}_n + \ri \zeta^{(2)}_n}{2}
\label{ab-eta-zeta}\nn\\
&& 
\{ \eta^{a}_n, \eta^{b}_m  \} = \{ \zeta^{a}_n, \zeta^{b}_m  \}
= 2 \delta^{ab} \delta_{nm}, ~
\{ \eta^{a}_n, \zeta^{b}_m  \} = 0~~~ \label{eta-zeta-algebra}
\eea
we find out that $H_0$ in (\ref{ham-rep-mod}) decouples into two 
identical 
Hamiltonians of the KM chain: 
\bea
&&~~~~H_0 = \sum_{a=1,2} H_{\rm KM} ^a, ~~[H_{\rm KM}^1, H_{\rm KM}^2] = 0\nn\\
&&H_{\rm KM}^a  
= \frac{\ri}{2}
\sum_{n=1}^N \left( t_+  \eta^a _n \zeta^a_{n+1} + t_- \eta^a_n \zeta^a_{n-1} 
-  t_{\perp} \eta^a_n \zeta^a_n \right), ~~~
~~\label{decoup1}
\eea
By the correspondence discussed in Appendix \ref{kitaev}, 
$H_{0}$ in (\ref{decoup1}) 
describes two copies of the XY spin-1/2 chains in a transverse
magnetic field:
\bea
&& H_0 = \sum_{a=1,2} H^a_{\rm XY}, \nn\\
&& H^a_{\rm XY} = - \sum_{n=1}^N \Big(
h \s^z _{a,n} 
+ J_x \s^x _{a,n} \s^x _{a,n+1} + J_y \s^y _{a,n} \s^y _{a,n+1}
\Big)\nn\\
&& \label{XY-sum}
\eea
or equivalently 
two decoupled 
$p$-wave superconducting chains: 
\bea
H_{0} &=& \sum_{a=1,2} H^a _{\rm 1DPS}, \nn\\
H^a _{\rm 1DPS} &=& \sum_n \Big[
- \mu_s \sum_n \left(f^{\dagger}_{a,n} f_{a,n} - 1/2\right)\nn\\
&+& t_s \left(  f^{\dagger}_{a,n} f_{a,n+1} + h.c.  \right)\nn\\
&+&  (\Delta_s/2) \left(  f^{\dagger}_{a,n} f^{\dagger}_{a,n+1} + h.c.  \right)
\Big] \label{1dps-ham1}
\eea
where $f^{\dagger}_{a,n} = (\zeta^a _n + \ri \eta^a _n)/2$ ~$(a=1,2)$. The parameters 
of the three above models are related by the formulas (\ref{param-rel}).
Notice that, according to
the definition (\ref{ab-eta-zeta}), the original fermionic operators $\alpha_n$ and $\beta_n$
mix up  Majorana species ($a=1,2$) belonging to \emph{different} KM chains to which the
Hamiltonian $H_0$ decouples.

According to the Jordan-Wigner 
equivalence (\ref{jw1}), the spin-fermion correspondence that relates
the models (\ref{decoup1}) and (\ref{XY-sum}) is highly nonlocal.
Each of the two KM models ($H_{\rm KM} ^{1,2}$) or XY spin chains ($H_{\rm XY}^{1,2}$)
is $\mathbb{Z}_2$-symmetric. In (\ref{decoup1})
this symmetry is realized as
the invariance 
under transformations
$\eta^a \to - \eta^a$, ~$\zeta^a \to - \zeta^a$, while
in (\ref{XY-sum}) it is the symmetry under $\pi$-rotations of the spin operators
$\s^x _{a,n} \to - \s^x _{a,n}$, $\s^y _{a,n} \to - \s^y _{a,n}$. 
However, the sum (\ref{decoup1}) possesses not only the discrete
$\mathbb{Z}_2 \otimes \mathbb{Z}_2$ symmetry but 
enjoys a larger, 
continuous SO(2) symmetry associated with global rotations of the Majorana vectors 
$\veta = (\eta^1, ~\eta^2)$ and $\vzeta = (\zeta^1, ~\zeta^2)$. 
This symmetry is nonlocally realized with respect to the spin model $H_0$ in (\ref{XY-sum}).
Its existence is consistent
with the U(1) symmetry of the original SSH ladder model and the related conservation of
the total fermion number.
\vskip 0.2cm

Having proven an exact equivalence between a noninteracting B-type dimerized ladder and a decoupled pair of two KM copies, which holds at arbitrary nonzero $t_{\perp}$,  we now specialize to the
vicinity of the 
Ising critical point $t_{\perp} = 2t_0 - m$, ~$|m|\ll t_0$. In this limit, one can
pass 
to a continuum description in which
$\eta^a _n \to \sqrt{2a_0} \eta^a (x)$, $\zeta^a _n \to \sqrt{2a_0} \zeta^a (x)$.
As a result
\begin{widetext}
\bea
H_{\rm KM}^a  \to \int \rd x~{\cal H}^a _{\rm KM} (x),~~
{\cal H}^a _{\rm KM} (x) =  \ri v \eta^a(x) \p_x \zeta^a(x) + \ri m \eta^a(x) \zeta^a(x)
\equiv ~ {\cal H}_{\rm QIC}[\eta^a,\zeta^a]
\label{KM-QIC}
\eea
\end{widetext}
where $v = \Delta a_0$. 
We see that, in the field-theoretical limit,
the 
KM chain reduces to a model of a massive Majorana fermion, which is nothing but
the continuum version
of a slightly off-critical Quantum Ising Chain (QIC) \cite{gnt, mussardo}. 
This fact is well known:  near an Ising transition the XY spin chain in a transverse magnetic field
is faithfully described by the Ising Field Theory \cite{ftnt1}.

A chiral rotation similar to that for the complex fermion field $\psi$, Eq.(\ref{chiral-rot}),
$\xi^{(a)}_{R,L} = (\eta^a \mp \zeta^a)/\sqrt{2}$,
leads to
\begin{widetext}
\bea
{H}_0 \to\int \rd x~ {\cal H}_0 (x), ~~~~
{\cal H}_0 (x)= \frac{\ri v}{2} \left( \vxi_R \cdot \p_x \vxi_R - \vxi_L \cdot \p_x \vxi_L  \right)
+ \ri m ~\vxi_R  \cdot \vxi_L, ~~~~{\vxi} = \left( \xi^{(1)}, \xi^{(2)}  \right)
\label{qic-sum}
\eea
\end{widetext}
This O(2)-invariant model of a two-component massive Majorana field
describes the staggered SSH
ladder near the U(1) criticality in the absence of interactions.
With a complex fermionic field $\chi (x)$ defined as
$\chi = (\xi^{(1)}  + \ri \xi^{(2)})/\sqrt{2}$ the model
(\ref{qic-sum}) is equivalent to a theory
of a free massive Dirac fermion in 1+1 dimensions, given by Eq.(\ref{dirac-canonical}).

Now we  consider the structure of interaction in the KM 
representation.
In terms of complex fermionic fields $\alpha_j$ and $\beta_j$ 
the fluctuation part of the interaction Hamiltonian (\ref{int}) takes the form 
\bea
&& H_{\rm int} = \sum_{j} \Big\{ U \delta \rho_{\alpha} (j)  \rho_{\beta} (j) \nn\\
&&+ V
\left[
\delta \rho_{\alpha} (j)  \rho_{\beta} (j+1) 
+ \delta \rho_{\alpha} (j)  \rho_{\beta} (j-1) 
\right]\Big\}
\label{int=alpha-beta}
\eea
where 
\bea 
&&\delta \rho_{\alpha} (j) = \alpha^{\dagger}_{j} \alpha_j - \frac{1}{2} = - \frac{\ri}{2} \eta^1_{j} \eta^2 _{j},
\nn\\
&&
\delta \rho_{\beta} (j) = \beta^{\dagger}_{j} \beta_j - \frac{1}{2} =  \frac{\ri}{2} 
\zeta^1 _{j} \zeta^2 _{j} 
\nn
\eea
So
\bea
H_{\rm int} &=&
\frac{1}{4}\sum_{j}
\Big\{  U \left( \eta_{1j} \zeta_{1j}  \right)\left( \eta_{2j} \zeta_{2j}  \right) \nn\\
&&+ V [ \left( \eta_{1j} \zeta_{1,j+1}  \right)\left( \eta_{2j} \zeta_{2,j+1}  \right)\nn\\
&&
+  \left(\eta_{1j} \zeta_{1,j-1}  \right)\left( \eta_{2j} \zeta_{2,j-1}  \right) ]
\Big\}
\label{UVV-maj}
\eea
Using the JW transformation from fermions to spin-1/2 variables (see Appendix \ref{kitaev}), 
it is interesting to reveal the spin-chain content of the O(2)-Majorana model with
interaction (\ref{UVV-maj}). We find that such fermionic model
is equivalent to the following
\emph{interacting} XY spin-ladder model:
\bea
H_{\rm XY}[\s_1, \s_2] &=& H_{\rm XY}^1 + 
H_{\rm XY}^2 
+ H'_{\rm XY} [\s_1, \s_2] \label{XY-AT-ham}
\eea
where $H_{\rm XY}^a ~(a=1,2)$ are given by 
Eq.(\ref{XY-sum}) and
\bea
H'_{XY} [\s_1, \s_2] &=&
\sum_j \Big[ U \s^z _{1,j} \s^z _{2,j} \nn\\
&+& V \sum_{\alpha=x,y} \s^{\alpha} _{1,j}\s^{\alpha} _{1,j+1}  \s^{\alpha} _{2,j}
\s^{\alpha} _{2,j+1} \Big]
\label{XY-AT-int}
\eea
is the interaction term. The Hamiltonian (\ref{XY-AT-ham}) can be regarded as an XY generalization of
the quantum Ashkin-Teller model, the latter describing a system of two quantum Ising chains near criticality, coupled
by a self-dual interaction \cite{dm,dsg}. This correspondence becomes relevant in the vicinity
of the Gaussian transition in the original SSH ladder model.
Indeed, at $t_{\perp} \sim 2t_0$ 
the interaction term (\ref{UVV-maj}) transforms to
\bea
H_{\rm int} &=& - \int \rd x~\Big\{ U  \eta_{1}(x) \zeta_{1}(x)  \eta_{2} (x) \zeta_{2}(x) \nn\\
&&+ V  \Big[ \eta_{1}(x) \zeta_{1}(x+a_0)  \eta_{2} (x) \zeta_{2}(x+a_0) \nn\\
&& + \eta_{1}(x) \zeta_{1}(x-a_0)  \eta_{2} (x) \zeta_{2}(x-a_0)\Big]
\Big\}
\label{int-major}
\eea
Taking the limit $a_0 \to 0$ in (\ref{int-major}) means keeping only the  part of
interaction which is marginal at the ultraviolet fixed point. All neglected terms
containing
derivatives of the fields, 
including those which describe Umklapp processes, are strongly irrelevant at the
Gaussian fixed point of the noninteracting model.
In this approximation one arrives at the true quantum Ashkin-Teller model
\bea
{\cal H}_{\rm AT} (x) &=& \sum_{a=1,2} \ri
\left[ v \eta^a (x) \p_x \zeta^a (x) + m \eta^a (x) \zeta^a (x) \right]\nn\\
&+& \lambda \eta^1 (x) \zeta^1 (x)  \eta^2 (x) \zeta^2 (x) 
\label{AT-true}
\eea
where $\lambda \sim g_+ \sim U+2V$.

The model (\ref{AT-true}) is equivalent to the bosonized Hamiltonian (\ref{DSG2}) 
without the Umklapp term.  
However, due to renormalizations caused by the marginal perturbation, the Luttinger-liquid parameter
may reach values $K \lesssim 1/2$, in which case Umklapp processes cannot be ignored.
It is clear that, to tackle effects caused by the (relevant) Umklapp processes
would be very hard, if possible at all, in the fermionic language, Eq.(\ref{AT-true}). On the contrary, the bosonization method 
reformulates the emerging problem in terms of the DSG model, which allows one to infer a valuable information
about the phase diagram and the new emerging criticalities.

\section{Nonlocal order parameters}\label{nonlocalOP}

The notion of nonlocal order in strongly correlated systems with a gapped spectrum
and unbroken continuous symmetry was originally associated with the Haldane spin-liquid
phase of the spin-1 chain.  den Nijs and Rommelse \cite{sop1} introduced a
string order parameter
\be
{\cal O}^{\alpha} = \lim_{|i-j|\to \infty} \Big\la S^{\alpha}_i
\exp\left( \ri \pi \sum_{k=j+1}^{i-1} S^{\alpha}_k  \right) S^{\alpha}_j \Big\ra,
\label{sop-NR}
\ee
$S^{\alpha}_i~(\alpha = x,y,z)$ being spin-1 operators, 
which takes a nonzero value in
the ground state. This was shown to be related to a spontaneous breakdown  of a hidden 
$\mathbb{Z}_2 \otimes \mathbb{Z}_2$ symmetry \cite{kom-tas}. The Affleck-Kennedy-Lieb-Tasaki
valence-bond state of a spin-1 chain \cite{AKLT} has revealed a deep connection between a
nontrivial topological order and  four-fold degenerate spin-1/2 boundary
states existing for an open chain.
Later on, nonlocal string order parameters were studied in 
various spin chains and ladders \cite{snt} to categorize massive phases of these objects according
to topologically distinct classes \cite{kim}.
In recent studies, string order parameters together with another nonlocal order parameter,
the parity operator which arose in the context of the Kitaev 1DPS model \cite{kitaev,kitaev-laumann}, were extensively studied to characterize
topologically trivial and nontrivial massive ground state phases of various
one-dimensional fermionic systems -- SSH and Kitaev chains and their quasi-1D analogs \cite{pollmann,bahri,catin}.

Nonlocal string-order and parity operators relevant to our discussion were considered earlier
for 1D lattice bosons \cite{giam2}. Below we show that, for one-dimensional fermions, a field-theoretical representation of these non-local operators remains the same.
Let us introduce the number of the original fermions
within the interval $1 \leq j \leq n$, measured from its average value 
\bea
\delta{\cal N}_n = {\cal N}_n - n = \sum_{j=1}^n \delta \rho_j = 
\sum_{j=1}^n \left[   \delta \rho_{\alpha}(j) +  \delta \rho_{\beta}(j) \right]
\nn
\eea
where $\delta\rho_j=\rho_j - 1$ is the fluctuation of the rung density.  We then define
the parity operator 
\bea
{\cal P}_n &=& e^{i\pi \delta {\cal N}_n}  
= (-1)^n \prod_{j=1}^n \left(  1 - 2 \alpha^{\dagger}_j \alpha_j \right)
\left(  1 - 2 \beta^{\dagger}_j \beta_j \right) \nn\\
&=&
\prod_{j=1}^n \left( \ri  \eta^{(1)}_j  \zeta^{(1)}_j \right)
\left( \ri  \eta^{(2)}_j  \zeta^{(2)}_j \right)
= P^{(1)}_n  P^{(2)}_n,
\label{parity-product}
\eea
The already discussed equivalence of the SSH ladder to two copies of the Kitaev chain
reveals the multiplicative structure of the operator ${\cal P}_n$: 
it is  a product of the parity operators of
the two copies of the Kitaev chain to which the SSH ladder Hamiltonian maps,
Eq.(\ref{decoup1}). A detailed analysis of the Majorana structure of nonlocal order parameters for an individual
Kitaev model can be found in Refs. \cite{bahri, chitov2} (see also Appendix \ref{kitaev}).

According to the bosonization rules, at $n\gg 1$ the local operator $\delta \rho_{n}$ transforms to
$({a_0}/{\pi}) \p_x \phi(x)$. Therefore, in the continuum limit
\bea
{\cal P}_n \to {\cal P} (x) &=&  \Re e :\exp \left[ \ri \sqrt{\pi} \int ^x \rd y~\p_y \phi(y)\right]:
\nn\\
&=& :\cos \sqrt{\pi} \phi(x):
\label{corr-parity}\eea
At the Gaussian fixed point
${\cal P} (x)$ is a primary field 
with scaling dimension $1/4$. Its nonlocal fermionic origin follows from the observation that it cannot
be expressed as a linear combination of fermionic mass bilinears (the Gaussian scaling dimension of the latter is 1). 
At infinite separation the two-point correlation function of local parity operators becomes
\[
\lim_{|x-y| \to \infty} \la  {\cal P} (x) {\cal P} (y) \ra
=  \la {P} \ra^2
\]
where $P =e^{i\pi {\cal N}}= P_1 P_2$ 
is the global parity operator, ${\cal N} = {\cal N}_N$ is the total particle number, $P_a~(a=1,2)$
are global parities  of the corresponding Kitaev chains (or related QIC models).

To obtain an equivalent representation of parity $\la P \ra$ in terms of the discrete (Ising)
variables, one can proceed either from the factorization
formula (\ref{parity-product}) and then use the results collected in Refs. \cite{bahri,chitov2},
or take advantage of the correspondence between a nearly critical staggered SSH ladder and 
the quantum Ashkin-Teller model for which main bosonization formulas are well known \cite{ZI, dsg}.
We will take the second route. Here
one should take into account the fact that, as
compared to the convention adopted in Ref.[\onlinecite{dsg}], in our case the sign of the Dirac (or Majorana) mass is inverted. Changing $m\to - m$ is equivalent to the duality transformation of the QIC model. With this circumstance in mind, one obtains
\be
P (x) \sim  \s_1 (x) \s_2 (x) \label{P-sigmas}
\ee
where $\s_j (x) ~(j=1,2)$ are local order parameters of the $j^{\rm th}$ Ising copy.

We now build up a string operator:
\bea
{\cal O}_S (n) &=& \exp\left( \ri \pi \sum_{j=1}^{n-1}\delta \rho_{j}   \right)
\delta \rho_{n} \equiv P_{n-1} \delta \rho_{n} \label{string-A}
\eea
In the continuum limit
\bea
{\cal O}_S (j) \to {\cal O} _S (x) =  \frac{a_0}{\pi}:\cos \sqrt{\pi} \phi(x-a_0):~\p_x \phi(x) 
\nn
\eea
Of interest is the string correlation function
\[
\lim_{|x-y| \to \infty} \la {\cal O} _S (x) {\cal O} _S (y) \ra = \la {\cal O} _S \ra^2
\]
In the conformal field theory of a massless Gaussian field  \cite{ginsparg, cft}, Eq.(\ref{masslessFB}),
the following operator product expansion can be derived
\bea
&&\p_x \phi(z, \bar{z}) :\cos \beta \phi(w,\bar{w}): \nn\\
&& =  \frac{\ri \beta}{4 \pi} \left( \frac{1}{z - w} - \frac{1}{\bar{z} - \bar{w}} \right)
: \sin \beta \phi (w, \bar{w}):  \label{o1} 
\eea
Here $z=v\tau + \ri x$, $\bz = v\tau - \ri x$ are complex variables,
$\tau$ being imaginary time. Setting $\tau = 0$ and substituting $\beta = \sqrt{\pi}$ and $z-w = \alpha$ (here $\alpha$ is the
short-distance cutoff of the bosonic theory),
up to a nonuniversal multiplicative constant we obtain
\bea
{\cal O} _S (x) \sim ~
 :\sin \sqrt{\pi} \phi(x):~ \sim \mu_1 (x) \mu_2 (x)
\label{string}
\eea
where $\mu_{1,2} (x)$ are the Ising disorder operators of the corresponding chains.
Referring for a general definition to chapter 9 of the book by Mussardo~\cite{mussardo}
here we only mention that the existence of the disorder operator in the QIC model
follows from the Kramers-Wannier diality between the ordered and disordered massive
ground-state phases that map to each other under sign reversal of the
Majorana mass. Physically, in the ordered Ising phase, the disorder operator $\mu (x)$ 
creates a kink at a point $x$ separating $\mathbb{Z}_2$-degenerate
states with opposite signs of the magnetization $\la \s (x)\ra$ at $x>0$ 
and $x<0$. Kink condensation
associated with the appearance of a nonzero average $\la \mu \ra$ 
leads to the onset 
of the disordered Ising phase.

When the marginal part of interaction is taken into account, the compactification radius of the scalar field gets changed and $\phi(x) \to \sqrt{K} \phi(x)$.  Accordingly, the parity and string
operators, (\ref{corr-parity}) and (\ref{string}), become 
\be
{\cal P} (x) \sim  :\cos \sqrt{\pi K} \phi(x):, ~{\cal O} _S (x) \sim 
 :\sin \sqrt{\pi K} \phi(x):
~~ \label{nonlocal-K}
\ee
Thus, the nonlocal fermionic operators of the staggered SSH ladder, parity and string order parameters, 
admit a local representation in terms of vertex operators of the scalar field $\phi(x)$.

\section{Phase diagram}\label{phase-dia}

We now turn to the low-energy effective bosonized model (\ref{DSG2}) and analyze the ground-state phase
diagram of the system in the vicinity of the critical point $t_{\perp} = 2t_0~(\gamma =0)$
as a function of the Luttinger-liquid parameter $K$, the deviation from criticality
($m \sim\gamma$) and the Umklapp coupling constant ($\tilde{g} \sim g_- \sim U-2V$).

\subsection{Case ${\bf {1}/{2} < K < 2}$}

This is a situation when 
$d_1 < 2, ~d_2 > 2$, so that the
Umklapp term in (\ref{DSG2}) is irrelevant  and the low-energy physics is described by the standard
quantum sine-Gordon model
\be
{\cal H} (x) = \frac{u}{2} \left[ \pi^2 (x) + \left(\p_x \phi(x)   \right)^2   \right]
+ \frac{m}{\pi \alpha} \cos \sqrt{4\pi K} \phi (x)
\label{sg-no-umkl}
\ee
Upon renormalization this model flows towards a strong-coupling fixed point characterized by a dynamically generated mass gap $M \sim |m|^{1/(2-K)}$.
 In the ground state, the field $\phi$ is locked
in one of the degenerate minima of the cosine potential:
\bea
(\phi)_{n} = \sqrt{\frac{\pi}{K}} \left[n+\frac{1}{2} \theta(m)\right], ~~n=0, \pm 1, \pm 2, \ldots
~~\label{vac-f}
\eea
where $\theta(x)$ is the Heaviside step function.
Since the separation of neighboring minima is $\Delta \phi = \sqrt{\pi/K}$, 
the fermionic number (\ref{Q-top}), associated with a topological kink of the SG model (\ref{sg-no-umkl}), is equal to $Q_F=1$.
This is the charge carried
by a massive Dirac fermion. The ground state of the system is 
insulating.

According to (\ref{Dbos}) and (\ref{vac-f}), at any $m \neq 0$ the ground state 
is characterized by
both the longitudinal and transverse explicit dimerization,
with averages $\la {\cal B}_{\parallel} \ra$ and $\la {\cal B}_{\perp}\ra$ 
nonzero and of opposite sign. 
At the critical point 
both bond-densities 
change their sign.
The system goes through a U(1),
Gaussian criticality to another massive phase. 
At $m=0$ 
the model displays the properties of a spinless Tomonaga-Luttinger liquid
characterized by the absence of single-fermion quasiparticles and power-law decay of
correlation functions with non-universal, $K$-dependent critical exponents (see for a review 
Refs.\cite{gnt, giam}).

The two massive phases with opposite signs of $m$ are dual to each other and
thermodynamically indistinguishable. However they differ in their topological properties.
The case $m<0$ corresponds to the disordered Ising phase while $m>0$ to the ordered phase.
In a single Kitaev 1DPS chain, the  Ising ordered (disordered) phases correspond to topologically nontrivial (trivial) phases of the superconductor. Then, according to the relations (\ref{P-sigmas})
and (\ref{string}),
we conclude that
\bea
m > 0:~~\la P \ra= 0,  ~~\la O_S \ra \neq 0 
&& ~~  \textrm{(topological  phase)}\nn\\
m < 0: ~~\la P \ra \neq 0, ~~\la O_S \ra = 0 
&& ~~  \textrm{(non-topological  phase)}\nn\\
\label{ave}
\eea
in full agreement with the different structure of the bosonic vacuum at $m>0$ and $m<0$ as
displayed by Eqs.(\ref{vac-f}). 
In the phases where they are nonzero, up to a nonuniversal coefficient
both parity and string order parameter scale with the bare mass $m$ as 
\bea
&& P(m) \sim \theta(- m) F(m), ~~
O_S (m) \sim \theta(m) F(m), \nn\\
&& ~~~~~~~~~F(m) \sim (|m|\alpha/u)^{K/4(2-K)}
\eea
On approaching the Gaussian criticality ($m \to 0$) both $P(m)$ and ${O}_S (m)$ vanish.

The above results for parity agree with the conclusions reached  by Kitaev and co-authors 
\cite{kitaev-laumann, FK}
who discussed topological properties of fermions in one dimension.
They argued that
in the topologically trivial phase of a 
1D $p$-wave superconductor 
the ground state has a certain parity. On the other hand,
in the topologically nontrivial phase
with two boundary Majorana zero modes, 
the nonlocally realized $\mathbb{Z}_2$ degeneracy of the vacuum always remains unbroken, 
and the average parity vanishes. We see that the bosonization treatment of
a pair of the Ising models to which the original SSH ladder maps
supports this conclusion.

The situation with the string order parameter is just the opposite.
From (\ref{ave}) it follows that
the operator ${\cal O} _S$ acquires a nonzero expectation value 
in the Ising ordered phase ($m>0$) and vanishes in the disordered Ising phase
($m<0$).
Thus, as expected,  the string order parameter is indicative of topological order in the model.

The insulating state at $m>0$ is topologically non-trivial. For the model (\ref{sg-no-umkl})
with open boundary conditions the spectrum contains boundary modes 
which transform to zero-energy midgap states in the thermodynamic limit ($L\to \infty$).
It is well-known \cite{JR} that each zero mode accumulates the fractional
charge $q_F = 1/2$. In bosonization language this fact can be understood as follows. A boundary of a 
finite
system, say  at $x=0$, is topologically equivalent to  a mass kink of the SG model (\ref{sg-no-umkl})
which 
separates the topological bulk phase ($x>0$) with $m > 0$
from the vacuum at $x<0$, the latter treated as a phase with $m \to -\infty$.
Following Jackiw and Rebby \cite{JR}, one then  replaces the the mass $m$ in (\ref{sg-no-umkl})  by a coordinate-dependent
function $m(x)$ with a solitonic profile:
$m(x) \to m > 0$ at $x \to \infty$, ~$m(x) \to -\infty$ at $x \to -\infty$.
The vacua corresponding to different signs of $m(x)$ have a relative shift 
$\Delta \phi = \sqrt{\pi/4K}$, which immediately leads to the fractional charge  $q_F = 1/2$
of the zero fermionic mode at the boundary, as opposed to the unit charge of the bulk fermionic excitations.
The bulk massive phase at $m<0$ is
topologically trivial: no boundary zero modes exist in this case.

It is worth noticing that,
in both massive phases, the link-parity symmetry ($P_L$) of the ground states 
\be
P_L : ~~n \to 1 - n, ~~\chi(x) \to \htau_2 \chi(-x), ~~~\phi(x) \to - \phi(-x)
\label{link-P}
\ee
excludes
the formation of a site-diagonal charge-density wave: $\la \rho_{\rm CDW} \ra = 0$.
Indeed, in the strong-coupling regime, for both sets of vacua (\ref{vac-f})
the average $\la  \sin \sqrt{4\pi K} \phi\ra$ vanishes.

Thus the properties of the system  at ${1}/{2} < K < 2$ are controlled by the magnitude
and sign of the Dirac mass $m$.

\subsection{Case ${\bf K < 1/2}$}

\begin{figure}[hbbp]
\centering
\includegraphics[width=3.4in]{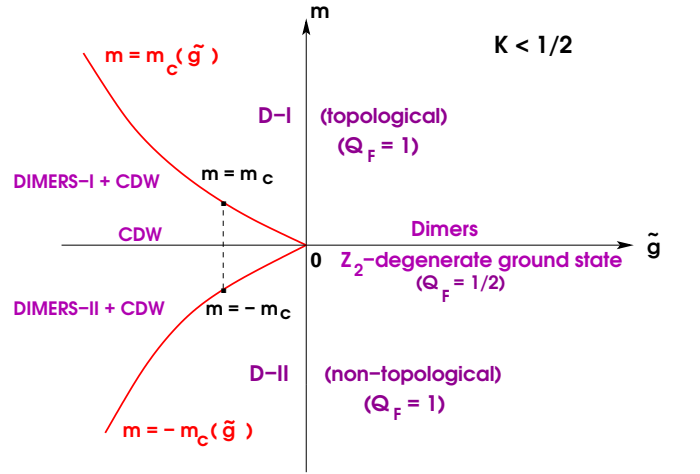}
\caption{\footnotesize Phase diagram of interacting staggered SSH ladder at $t_{\perp} \sim 2t_0$, $K<1/2$.
D-I and D-II denote non-degenerate massive phases with opposite signs
of average dimerization. At $m=0$ the ground state is two-fold degenerate and is spontaneously
dimerized if 
$\tilde{g}>0$ or has a site-diagonal charge-density wave if $\tilde{g}<0$. $Q_F$ indicates the
fermionic charge of elementary excitations. $m = \pm m_c (\tilde{g})$ are critical lines belonging
to the Ising universality class. A similar phase diagram for $t_{\perp} \sim - 2t_0$
is obtained from the present one by a mirror reflection with respect to the horizontal axis.}
\label{dsg-regime}
\end{figure}

A more complicated and interesting picture emerges
when both perturbations 
in the DSG model (\ref{DSG2})
are relevant: 
$d_1 < d_2 < 2 $. The phase diagram of the system at $K<1/2$ is schematically depicted in Fig.\ref{dsg-regime}.

Suppose that the noninteracting ladder is at the Gaussian critical point $m=0$.
When Umklapp
processes are taken into account, the effective low-energy theory is described by a sine-Gordon model
but with a different cosine perturbation:
\bea
{\cal H} (x) &=& \frac{u}{2} \left[ \pi^2 (x) + \left(\p_x \phi(x)   \right)^2   \right]\nn\\
&-& \frac{\tilde{g}}{2(2\pi \alpha)^2} \cos \sqrt{16\pi K} \phi (x)
\label{new-SG}
\eea
At $K\leq 1/2$ the dynamically generated mass gap scales as $M_{\tilde{g}} \sim |\tilde{g}|^{1/(2-4K)}$.
In the infrared limit the field $\phi$ gets locked in one of the minima
\bea
(\phi)_n = \frac{1}{2} \sqrt{\frac{\pi}{K}} \left[(n+\frac{1}{2} \theta (-\tilde{g})\right], 
~~~n = 0, \pm 1, \pm 2, \cdots \label{set}
\eea
Since $\Delta \phi = (\phi)_{n+1} - (\phi)_n =  \sqrt{{\pi}/{4K}}$, one concludes that the
the fermionic number carried by  a quantum soliton of the SG model (\ref{new-SG})
is fractional, $Q_F = 1/2$. 
The critical point  $\tilde{g} = 0$ 
separates two massive phases with different physical properties. At $\tilde{g} > 0$
$\la \cos \sqrt{4\pi K} \phi \ra \neq 0$, $\la \sin \sqrt{4\pi K} \phi \ra = 0$, so there exists
a bond-density wave in ground state of the system (dimerization). 
On the other hand, at
$\tilde{g} < 0$ $\la \cos \sqrt{4\pi K} \phi \ra = 0$, $\la \sin \sqrt{4\pi K} \phi \ra \neq 0 $,
indicating the onset of a site-diagonal charge-density wave. 
Close to the critical point both the  dimerization $\la {\cal B}_{\parallel, \perp} \ra$ and the staggered density $\la \rho_{\rm CDW} \ra$ (at $\tilde{g}>0$ and $\tilde{g} <0$, respectively) scale as 
\bea
|\la {\cal B}_{\parallel, \perp} \ra| \sim |\la {\rho}_{\rm st} \ra|
\sim \left( |\tilde{g}|/{u} \right)^{K/2(1-2K)}
\eea
A similar phase diagram has been discussed by Haldane \cite{haldane-xxz} 
for the XXZ spin-1/2 Heisenberg
antiferromagnetic chain with competing interactions (in the latter case, by the Jordan-Wigner
correspodence the Neel order translates to the CDW one).

The two massive phases have different
symmetry properties.  The dimerized phase is link-parity symmetric while site parity $P_S$
\be
P_S: ~n \to - n, ~\chi(x) \to \htau_1 \chi(-x), ~\phi(x) \to \frac{\sqrt{\pi}}{2} - \phi(-x)
\label{parity-S}
\ee
is \emph{spontaneously}
broken. For the CDW phase the situation is just the opposite.

 In both phases the ground state 
is doubly degenerate. 
This follows from the fact that
the corresponding order parameters, $\la \cos \sqrt{4\pi K} \phi \ra$ and
$\la \sin \sqrt{4\pi K} \phi \ra $, have opposite signs for even and odd values of the integer
$n$ which in (\ref{set}) 
labels different degenerate vacua.  The aforementioned 
quantum solitons of the SG model (\ref{new-SG}) are the kinks interpolating between
the degenerate vacua. Fermions as stable quasiparticles are absent in the spectrum.
In the spontaneously dimerized phase ($\tilde{g} > 0$) different degenerate vacua have different topological properties. It follows from (\ref{set}) that at $\tilde{g} > 0$ the average parity and string order parameters are proportional to $\cos (\pi n /2)$ and $\sin (\pi n /2)$, respectively, implying that
only one of the two degenerate dimerized phases is topological (namely the one with $n$ odd for which
$\la P \ra=0$, $\la {O}_S \ra \neq 0$) while the other is not ($\la P \ra \neq 0$, $\la O_S\ra = 0$).
The CDW phase at $m=0$, $\tilde{g} < 0$  is a "topologically mixed" phase. As seen from (\ref{set}),
in any of the degenerate CDW vacua both $P$ and $O_S$ have nonzero vacuum expectation
values. We will return to this point in the sequel.

Consider now small deviations from the critical point, $m \neq 0$, keeping $K < 1/2$. Then the mass term is important, and one has to proceed from the DSG model (\ref{DSG2}) with both perturbations present. Apparently, the 
term $m \cos \sqrt{4 \pi K} \phi$ is the most relevant perturbation. However
when 
$m$ is small enough,  the interplay of the two perturbations determines the nature of the infrared fixed point. The crossover to the new low-energy regime will occur when
the mass gaps that would be generated separately by each of the two perturbations in Eq.(\ref{DSG2})
are of the same order: 
\be
\left( |m|\alpha/{u}  \right)^{\frac{1}{2-K}} \sim
\left( |\tilde{g}|/{u}  \right)^{\frac{1}{2(1-K)}} \label{same-order}
\ee

If $\tilde{g} > 0$, the vacua (\ref{vac-f}) 
are odd and even subsets
of the set  
(\ref{set}). The $m$-perturbation lifts the degeneracy between the two sublattices
of the potential $\tilde{g} \cos \sqrt{16 \pi K} \phi$ and leads to the period doubling
in $\phi$-space. The new period is that of the potential $m\cos \sqrt{4\pi K}\phi$.
As a result, two kinks with fractional charges $Q_F =1/2$ confine to produce a bound state 
which is equivalent to recovery of the fundamental fermion with
an integer charge $Q_F = 1$.  Thus at $\tilde{g}>0$
main properties of the massive phase of the DSG model (\ref{DSG2}), including
the quantum numbers of topologically stable excitations,  are essentially the same as
in the absence of Umklapp processes. Nevertheless, as $m \to 0$, the system does go through
a critical point: just due to Umklapp processes the spectral mass gap $M(m)$ undergoes a discontinuity:
\[
\lim_{m \to \pm 0} M (m) = \pm |M_{\tilde{g}}|.
\]

The situation qualitatively 
changes when $\tilde{g} < 0$. Now the sets of fields 
(\ref{vac-f}) 
that minimize the potential $m\cos \sqrt{4\pi K}\phi$ do not minimize 
$|\tilde{g}| \cos \sqrt{16 \pi K} \phi$.  The DSG potential in this case undergoes 
a topological transition. 
Let us illustrate this semi-classically \cite{dm}.
Consider the potential ${\cal U} (\varphi) = \mu \cos \varphi 
+ g \cos 2\varphi$, where $g \sim - \tilde{g} > 0$,  $\mu \sim m$ and
$\varphi = \sqrt{4\pi K} \phi$.
At $\mu<4g$ the potential displays a set of degenerate minima located at
\be
\varphi = \pm \left({\pi}/{2} + \eta_0 \right), ~~~{\rm mod}~2\pi
\label{dw-minima}
\ee
where $\sin \eta_0 = \mu/4g$. These minima are assembled in
a sequence of local 
double-well potentials. At $\mu/4g \to \pm 1$ the two minima of each  
double-well potential merge 
($\eta_0 \to \pi/2~{\rm sgn}~\mu$), and ${\cal U}$ becomes $2\pi$-periodic, with minima located either at
$\varphi = (2n+1)\pi$ at $\mu > 0$ or $\varphi = 2\pi n$ at $\mu < 0$. The conditions $\mu/4g = \pm 1$ provide 
classical values of two symmetric critical points. 
The double-well potential structure of ${\cal U}$
implies (in the Ginzburg-Landau sense) that the transition should belong to the Ising universality class.
A quantum estimate of the Ising critical lines follows from the relation (\ref{same-order}):
up to a non-universal multiplicative numerical constant these lines are determined by the equations
\be
m = \pm m_c (\tilde{g}), ~~~ m_c (\tilde{g}) \sim \frac{u}{\alpha}
\left( \frac{|\tilde{g}|}{u}  \right)^{\frac{2-K}{2(1-K)}}
\label{ising-crit-curves}
\ee

Thus, at a given $\tilde{g} < 0$, as $|m|$ is increased from $m=0$, there exists critical values 
$m = \pm m_c$ at which the system undergoes a quantum Ising transition (see Fig.\ref{dsg-regime}). 
A precise value of the phase shift $\eta_0$ in terms of the parameters $m$, $\tilde{g}$ and $K$
is unknown. However for us only two limiting values of $\eta_0$ are important:
$\eta_0 \to 0$ at $m \to 0$ and $\eta_0 = \pm \pi/2$ at $m \to \pm m_c$.
The inner region
bordered by the two Ising critical lines represents a
mixed phase in which dimerization $\la B_{\parallel,\perp} \ra \neq 0$ coexists with the site diagonal CDW,
$\la \rho_{CDW} \ra \neq 0$
(see Fig.\ref{dsg-regime}).
The average dimerization $\la B \ra$ changes its sign at $m=0$ but remains finite at the critical lines
$m=\pm m_c$.  The CDW order parameter reaches its maximum at $m=0$, vanishes
on approaching the critical lines and remains zero in the
regions  $|m|>m_c$ where the $P_L$ symmetry of the ground state is recovered.

The behavior of the system in the vicinity of each of the two Ising critical points $m= \pm m_c$ are described
in terms of an effective Ising field theory. Necessary details can be found in Ref. \cite{dsg}.
Adopting the ultraviolet-infrared transmutation of the physical fields found in \cite{dsg},
we read off the singular parts of the average dimerization and CDW in the regions
$|m \mp m_c| \ll m_c$:
\bea
&& \la B_{\parallel,\perp} \ra_{m} - \la B_{\parallel,\perp} \ra_{m_c}
\sim \pm \left( \frac{m\mp m_c}{m_c} \right)\ln \frac{m_c}{|m \mp m_c|}, \nn\\
&& \la \rho_{\rm CDW} \ra_m \sim \theta (m_c - |m|)\Big|   \frac{|m \mp m_c|}{m_c}\Big|^{1/8}, ~~~~
\label{B-CDW}
\eea

According to (\ref{dw-minima}), in the mixed phase ($m<m_c$) the fractional soliton of the SG
model (\ref{new-SG}) with $Q_F = 1/2$ splits into two topological kinks carrying
charges  
\be
Q_F ^{\pm}  = \frac{1}{2} \mp \frac{\eta_0}{\pi} \label{frac-charges1}
\ee
The existence of excitations in the mixed phase ($|m|<m_c$), carrying fermionic numbers that
continuously depend on the parameter $\eta_0$, follows from the spontaneous breakdown of charge conjugation symmetry $\mathbb{C}$ caused by the onset of a CDW: 
\bea
\mathbb{C}: ~~&& R(x) \to R^{\dagger}(x), ~L(x) \to L^{\dagger} (x),~
\Phi(x) \to - \Phi(x)\nn\\
&& ~~~~\mathbb{C} Q \mathbb{C}^{-1} = - Q, ~~~ \mathbb{C} \rho_{\rm st} \mathbb{C}^{-1} = - \rho_{\rm st}\label{charge-conj}
\eea
The polymer {\sl cis}-polyacetylene is a example of this kind \cite{ns}.
At the Ising critical points ($\eta_0 \to \pm \pi/2$) the two kinks merge and
the standard classification of integer topological
quantum numbers is recovered: $Q_F = 0,1$. 
It is just the singlet kink which loses its topological charge and becomes massless at 
$m=\pm m_c$. $Q_F = 1$ is the standard fermionic number of the explicitly
dimerized phases, D-I and D-II in Fig.\ref{dsg-regime}.

Some information on
topological properties of the ladder in the mixed phase of the phase diagram
($\tilde{g}<0$, $|m < m_c|$)
can be extracted from semi-classical estimates.
Using the location of the minima of the DSG potential, given by
(\ref{dw-minima}), for average parity an string order parameter, Eqs. (\ref{nonlocal-K}),
we obtain 

\bea
\la P \ra \sim \cos (\varphi/2) = \frac{1}{\sqrt{2}} \Big| \cos \frac{\eta_0}{2} - \sin \frac{\eta_0}{2}  
\Big|
\label{P-last}\\
\la O_S \ra \sim  \sin (\varphi/2) = \frac{1}{\sqrt{2}} \Big| \cos \frac{\eta_0}{2} + \sin \frac{\eta_0}{2}  \Big|
\label{S-last}
\eea
As already mentioned, at $m = 0$ ($\eta_0 = 0$) both $\la P \ra$ and $\la O_S \ra$
are nonzero.
When $m \to m_c - 0$ ($\eta_0 \to \pi/2$), $\la P \ra$ vanishes while
$\la O_S \ra$ is finite. This is consistent with the fact that the phase D-I ($m>m_c$) is topologically nontrivial.
On the other hand, at $m \to -m_c + 0$ ($\eta_0 \to -\pi/2$), $\la P \ra$ remains finite while 
$\la O_S \ra \to 0$,
indicating that at the lower Ising critical point
the system enters a topologically trivial phase D-II ($m < - m_c$).

\section{Conclusion}

In this paper, we have studied the ground-state phase diagram of the interacting staggered two-chain
SSH ladder in the vicinity of Gaussian critical point ($t_{\perp} \sim 2t_0$).
We have derived a fully bozonized effective field-theoretical model to treat
correlations effects in a non-perturbative way. We have shown that such model has the structure
of the double-frequency sine-Gordon (DSG) model \cite{dm,dsg}, Eq. (\ref{DSG2}),
characterized by the existence of
two perturbations at the Gaussian fixed point: the deviation from criticality 
parameterized in terms of a "Dirac mass"
$m \sim 2t_0 - t_{\perp}, ~|m|\ll 2t_0$,
and four-fermion Umklapp scattering
processes with amplitude $\tilde{g}$. The effects of forward scattering of the particles are 
phenomenologically incorporated
into a Luttinger-liquid parameter $K$ which  varies in a broad interval
including the region where both perturbations are relevant.

Massive phases with an explicitly or spontaneously broken symmetry  have been identified by 
inspecting order parameters described by expectation values of local fermionic fields in the
bosonic reprsentation. 
The structure of the nonlocal operators, parity and string order parameter, which identify
topologically non-trivial phases,
has been completely clarified as a result of a proof that a noninteracting fermionic staggered SSH ladder can be exactly
mapped onto a O(2)-symmetric model of two decoupled Kitaev-Majorana chains (or two 1D $p$-wave
superconductors). In the vicinity of the Gaussian fixed point, an interacting staggered
SSH ladder is equivalent to an Ashkin-Teller-like system of two coupled quantum Ising chains
with a non-locally realized O(2) symmetry. 
This equivalence made it possible to show that
topological order in the SSH ladder is related to broken-symmetry phases of the associated quantum spin-chain degrees of freedom.

At a relatively weak interaction ($1/2 < K < 2$) Umklapp scattering 
plays a subleading role, so that the ground state properties
of the model are dominantly controlled by the magnitude and sign of the Dirac mass $m$.
At $m=0$,  the ground state represents a Tomonaga-Luttinger liquid,
but at $m\neq 0$ it is 
explicitly dimerized and insulating.  Only the $m>0$ massive phase,
which is thermodynamically indistinguishable from its $m<0$ counterpart, is topological. 
At a stronger and longer-range interaction ($K<1/2$) both the mass and Umklapp perturbations are
relevant, and their interplay results in
the ground state phase diagram shown in Fig.\ref{dsg-regime}. At $m=0$ and any nonzero
$\tilde{g}$ the Tomonaga-Luttinger liquid becomes unstable under a transition to a spontaneously
dimerized state ($\tilde{g} > 0$) or a site-diagonal CDW  ($\tilde{g}<0$). Elementary excitations are
quantum solitons carrying fractional charge $Q_F = 1/2$. At $\tilde{g} > 0$
only one of the two degenerate dimerized phases is topological, whereas at $\tilde{g}<0$
the CDW phase is a "topologically mixed" phase with the both average parity and string order parameter
nonzero.

We have shown that in our model, depending on the sign of $\tilde{g}$, both scenarios of the DSG model
are realized: kink confinement
and  Ising quantum transitions.  At $\tilde{g} > 0$ the mass term lifts the degeneracy between the two spontaneously dimerized states and leads to confinement of two fractionally charged excitations, thus
resulting in the recovery of the fundamental fermion with a unit charge $Q_F =1$.
At $\tilde{g} < 0$ the phase diagram acquires Ashkin-Teller-like features. 
The Gaussian critical point splits into
two symmetric Ising critical lines $m = \pm m_c (\tilde{g},K)$, 
$m_c (\tilde{g},K) \sim |\tilde{g}|^{\frac{2-K}{2(1-K)}}$. 
These two lines sandwich
a mixed massive phase in which dimerization coexists with a site-diagonal CDW.
In this phase charge conjugation symmetry  is spontaneously broken and, consequently,  
the fermionic number $Q_F$ is not quantized 
in units 1/2. Elementary bulk excitations in the mixed phase are 
represented by
two types of topological solitons carrying different fermionic charges, which continuously interpolate between the values $Q_F =0$ and $1$.
This phase has also mixed topological properties with continuously varying parity and string order parameters. It would be very interesting to investigate the structure and spectrum of midgap
edge states in such mixed phase.

Cold atom setups are excellent candidates to realize a staggered dimerized ladder with a control of
its main parameters. Of particular interest and importance are topological properties
of this and other quasi-1D systems.
There has been a significant recent progress in developing novel experimental techniques using
optical miscoscopy, aimed
at observation of edge
states at interfaces, separating topologically distinct phases of 1D ultracold atomic systems
\cite{cheon, meier, rosch}. Remarkably, Ref. [\onlinecite{rosch}] reports on an experimental realization
of a Dirac model with an inhomogeneous mass term, directly related to an inhomogeneous
SSH chain.
We hope that the new methodology will make it soon possible to study 
topological excitations, including edge modes,
in multi-chain SSH setups, so that the results of this paper might potentially be 
relevant to future experimental studies.

The approach developed in this paper for a two-chain dimerized ladder can be straightforwardly generalized
to a larger number of chains. This would lead to a possibility to study correlation effects and topological properties of systems displaying quantum criticalities with non-Abelian symmetry groups.
This and related questions are presently under investigation.

\acknowledgements

I am grateful to Dionys Baeriswyl for bringing my attention to his earlier
works on SSH ladders \cite{BM1,BM2} and for interesting discussions.
I would also like to thank Marcello Dalmonte and Pierre Fromholz 
for their interest in this work and helpful comments,
and Rozario Fazio for his permanent support. 
I thank Mikheil Tsitsishvili
for his cooperation in studying the properties of the staggered SSH ladder in the massive incommensurate
regime. The support from the
Shota Rustaveli National Science Foundation of Georgia,
SRNSF, grant number FR-19-11872, is gratefully acknowledged.

\appendix

\section{Bosonization dictionary}\label{bos}

Here we provide some technical details related to the bozonization method \cite{gnt},  used
in the main text.
In 1+1 dimensions free massless fermions (case $m=0$ in Eq.(\ref{dirac-canonical}))
are equivalent to free massless bosons:
\bea
&&{\cal H}^{(0)}_B = \frac{v_0}{2} \left[ \left( \p_x \Theta  \right)^2 + \left(\p_x \Phi
\right)^2   \right] \nn\\
&&= v_0  \left[ \left( \p_x \varphi_R  \right)^2 + \left( \p_x \varphi_L \right)^2   \right]
= \pi v_0 \left( J^2 _R + J^2 _L  \right)~~
\label{masslessFB}
\eea
Here $\Phi (x) = \varphi_R (x) + \varphi_L (x)$ is a massless scalar field,
$\Theta (x) = - \varphi_R (x) + \varphi_L (x)$ is dual counterpart. The chiral currents
$J_{R,L} (x)$  are expressed in terms of the chiral bosonic field
$\varphi_{R,L} (x)$
\bea
J_R (x) = \frac{1}{\sqrt{\pi}} \p_x \varphi_R (x), ~~
J_L (x) = \frac{1}{\sqrt{\pi}} \p_x \varphi_L (x)~~\label{curr-fields}
\eea
and satisfy the U(1) 
Kac-Moody algebra \cite{gnt}
\bea
 [J_{R/L} (x), J_{R/L} (x')] &=& \pm \frac{\ri}{2\pi} \delta' (x-x') \nn\\
~~~~[J_R (x), J_L (x')] &=& 0
\label{kac}
\eea
Adding to (\ref{masslessFB}) the part of ${\cal H}_{\rm int}$ quadratic in the
currents $J_{R,L}$, we define the Gaussian part of the equivalent bosonic model:
\bea
&& {\cal H}_{\rm Gauss} 
= \frac{u}{2} \left( 1 - \frac{\lambda}{2\pi u}  \right) \Pi^2 
+  \frac{u}{2} \left( 1 + \frac{\lambda}{2\pi u}  \right) \left( \p_x \Phi  \right)^2\nn\\
&& \label{gaussian}
\eea
where 
$u = v_0 \left( 1 + {\lambda}/{2\pi v_0}   \right)$ is the renormalized velocity.
Here $\Pi (x) = \p_x \Theta (x)$ is the momentum conjugate to the field $\Phi(x)$. The
current algebra (\ref{kac}) ensures the canonical commutation relation
$
[\Phi(x), ~\Pi(x')] = \ri \delta(x-x'). 
$

The fermionic mass bilinears acquire the following bosonic representation
\bea
R^{\dagger} L \to
- \frac{\ri}{2\pi \alpha} e^{- i \sqrt{4\pi} \Phi}, ~~~
L^{\dagger} R \to \frac{\ri}{2\pi \alpha} e^{i \sqrt{4\pi} \Phi}~~~
\label{bilin1}
\eea
where $\alpha$ is the ultraviolet cutoff of the bosonic theory.
In particular
\bea
\chi^{\dagger} \htau_2 \chi &=& -\ri \left( R^{\dagger} L - h.c.
\right) \to - \frac{1}{\pi\alpha} \cos \sqrt{4\pi} \Phi \label{bilinear1}\\
\chi^{\dagger} \htau_1 \chi &=&  R^{\dagger} L + h.c.
~\to~ - \frac{1}{\pi\alpha} \sin \sqrt{4\pi} \Phi \label{bilinear2}\eea
Using point splitting  one bosonizes the Umklapp operator
\bea
{\cal O}_{\rm umkl} (x) &=& \left( R^{\dagger} L  \right)^2 _{x,x+\alpha}
+ \left( L^{\dagger} R  \right)^2 _{x,x+\alpha}\nn\\
&& - \frac{1}{2(\pi \alpha)^2} \cos \sqrt{16 \pi} \Phi(x)
\label{umkl-bos}
\eea
In the massless case ($m=0$) the effective bosonic model takes the form:
\bea
&&{\cal H}_B = {\cal H}_{\rm Gauss} + {g} {\cal O}_{\rm umkl}
= \frac{u}{2} \left[ K \Pi^2 + K^{-1} \left( \p_x \Phi  \right)^2   \right]\nn\\
&&~~~~~~~~~~~~- \frac{g}{2(\pi\alpha)^2} \cos \sqrt{16 \pi} \Phi
\label{ham-bos-total}
\eea
At a weak marginal coupling $\lambda$  the parameter $K$
is given by an expansion
\be
K = 1 - \frac{\lambda}{\pi u} + O(\lambda^2) \label{K}
\ee
in which only the $O(\lambda)$-term is universal. 
Generally, the Luttinger-liquid parameter $K$ decreases with increasing short-range repulsion, 
but for $K$
to reach arbitrarily small values longer range interaction is required \cite{giam}.

Rescaling the field and momentum
\[
\Phi (x) = \sqrt{K} \phi (x), ~~~~\Pi (x) = (1/\sqrt{K}) \pi (x)
\]
one rewrites (\ref{ham-bos-total}) as
\bea
{\cal H}_B (x) &=& \frac{u}{2} \left[ \pi^2 (x) + \left(\p_x \phi(x)   \right)^2   \right]\nn\\
&-& \frac{g}{2(\pi\alpha)^2} \cos \sqrt{16 \pi K} \phi~~~
\label{SG}
\eea
Eq.(\ref{SG}) is a quantum sine-Gordon (SG) model which is well-known to describe a 1D system
of spinless fermions with a nearest-neighbor density-density interaction. Equivalently,
such model describes scaling properties of the XXZ spin-1/2 chain \cite{gnt}.

At $m\neq 0$ one uses (\ref{bilinear1}) one bosonizes the Dirac mass term, in which case
the effective bosonic theory transforms to double-frequency sine-Gordon (DSG) 
model \cite{dm,dsg}:
\bea
{\cal H}_{\rm DSG} &=&  \frac{u}{2} \left[ \pi^2 (x) + \left(\p_x \phi(x)   \right)^2   \right]
+ \frac{m}{\pi \alpha} \cos \sqrt{4\pi K} \phi \nn\\
&-& \frac{g}{2(\pi\alpha)^2} \cos \sqrt{16 \pi K} \phi
\label{DSG-appen}
\eea

\section{Kitaev-Majorana chain and related models} \label{kitaev}

In this Appendix we collect known facts about the Kitaev-Majorana (KM) chain \cite{kitaev}
and its equivalent representations which are used in the bulk of this paper.
The KM chain is defined on a lattice with $N$ lattice sites in terms
of a pair of Majorana lattice fields, $\eta_n$ and $\zeta_n$:
\bea
&&H_{\rm KM}[\eta,\zeta] =\nn\\
&& \ri \sum_{n=1}^N \left(
- h \eta_n \zeta_n + J_x \eta_n \zeta_{n+1} - J_y
\zeta_n \eta_{n+1} \right)~~~
\label{S-etazeta}
\eea
In special cases $J_x \neq 0, ~J_y = 0$ or
$J_x = 0, ~J_y \neq 0$ the Hamiltonian 
  $H_{\rm {\sf KM}}$ reduces to a Quantum
Ising Chain (QIC) in the Majorana representation. 
In the general case $J_x \neq 0, ~J_y \neq 0$
the model (\ref{S-etazeta}) constitutes a Majorana representation of the
spin-1/2 XY chain in a transverse magnetic field 
\bea
H_{\rm XY}  = - h \sum_n \s^z _n - \sum_n \left(
J_x \s^x _n \s^x _{n+1} + J_y \s^y _n \s^y _{n+1} \right)
~~~~\label{XY+h}
\eea
The spin-chain model (\ref{XY+h}) is in turn equivalent to
the Kitaev toy model of a 1D $p$-wave superconductor (1DPS):
\bea
&&H_{\rm 1DPS} = - \mu_s \sum_n \left(f^{\dagger}_n f_n - \frac{1}{2}\right)
+ t_s \sum_n \left(  f^{\dagger}_n f_{n+1} + h.c.  \right)\nn\\
&& ~~~~~+~ (1/2)\Delta_s \sum_n \left(  f^{\dagger}_n f^{\dagger}_{n+1} + h.c.  \right)
\label{1dps-ham}
\eea
The equivalence of the three models, (\ref{S-etazeta}), (\ref{XY+h}) and
(\ref{1dps-ham}) is established in two steps.
First, by the 
Jordan-Wigner (JW) correspondence 
\bea
\s^z _n = 2f^{\dagger}_n f_n - 1, ~~~
\s^+ _n = 2 (-1)^n f^{\dagger}_n e^{i\pi \sum_{j=1}^{n-1} f^{\dagger}_j f_j}
~~~~\label{jw1}
\eea
$H_{\rm XY}$ is mapped onto $H_{\rm 1DPS}$.
The parameters of
the two  models are related as
\bea
\mu_s = 2h, ~~t_s = J_x + J_y, ~~\Delta_s = 2 (J_x -J_y)
\label{param-rel}
\eea
Secondly, splitting each complex fermion into a pair of Majorans fermions   \cite{kitaev}
\bea
&& ~~~~
~~f^{\dagger}_n =  (\zeta_n + \ri \eta_n)/2,~~~\zeta^{\dagger}_n = \zeta_n, ~~~
\eta^{\dagger}_n = \eta_n~~\nn\\
 \label{zeta-eta1} &&\{ \zeta_n , ~\zeta_m\} = \{ \eta_n , ~\eta_m\} = 2\delta_{nm},~~~~
\{ \zeta_n , ~\eta_m\} = 0 ~~~~~\label{zeta-eta} 
\eea
one transforms 
$H_{\rm 1DPS}$ to $H_{\rm KM}~[\eta,\zeta]$.

One can now build Majorana string operators \cite{bahri,chitov2}. Consider two-spin correlation functions for the XY spin model (\ref{XY+h}):
$\Gamma_x (1,n) = \la  \s^x _1 \s^x _n \ra$ and $\Gamma_y (1,n) = \la  \s^y _1 \s^y _n \ra$.
Spin ordering in the $x$ or $y$ directions of spin space is determined by
the asymptotical behavior of these correlation functions in the limit $n \to \infty$:
\be
\lim_{n\to \infty} \Gamma_x (1,n) = \la \s^x \ra^2, ~~~~
\lim_{n\to \infty} \Gamma_y (1,n) = \la \s^y \ra^2
\label{xy-s}
\ee
Since  for any fixed spin projection ($\alpha=x,y,z$)
$
\s^\alpha _1 \s^\alpha _n = \left( \s^\alpha _1 \s^\alpha _2  \right)  
\left( \s^\alpha _2 \s^\alpha _3  \right) 
\cdots  \left( \s^\alpha _{n-1} \s^\alpha _n  \right),
$
from (\ref{jw1}) one deduces that
\bea
\s^x _n \s^x _{n+1}  = - \ri \eta_n \zeta_{n+1}, ~~~~~~~~
\s^y _n \s^y _{n+1} = \ri \zeta_n \eta_{n+1}
\label{relations}
\eea
and finds out that the correlation functions $\Gamma_{x,y} (1,n)$,
local in spin variables, are non-local in terms of the fermions, in which case they represent
\emph{string order parameters}:
\bea
\Gamma_x (1,n) &\equiv& {\cal O}_x (n) = \prod_{j=1}^{n-1} (-\ri \eta_j \zeta_{j+1}), \nn\\ 
\Gamma_y (1,n) &\equiv& {\cal O}_y (n) = \prod_{j=1}^{n-1} (\ri \zeta_j \eta_{j+1}) \label{xy-eta-zeta}
\eea
In the region $|\mu_s| < 2t_s$ $~(|h| < J_x + J_y)$, depending on the sign of $\Delta_s = 2(J_x - J_y)$, 
either $\la \s^x\ra \neq 0$, $\la \s^y\ra = 0$ or $\la \s^x\ra = 0$, $\la \s^y\ra \neq 0$.
Consequently, either the string order parameter
${\cal O}_x (n)$ acquires a nonzero expectation value at $n \to \infty$ if $\Delta_s > 0$,
or ${\cal O}_y (n)$ if $\Delta_s < 0$ \cite{chitov2}. At $|\mu_s| > 2t_s$ $~(|h| > J_x + J_y)$ $\la \s^x \ra = \la \s^y \ra = 0$
and string order is absent. The direct calculations of the topological invariant 
\cite{kitaev, alicea} define the region  $|\mu_s| < 2t_s$ where the massive phase
of the  Kitaev's 1DPS model (\ref{1dps-ham}) is topologically non-trivial.
The same conclusion is reached when the string order parameters ${\cal O}_{x,y} (n)$
are analized in the limit $n\to \infty$. This fact illustrates the efficiency of the string order
in the studies of topological phases of 1D Fermi systems.

\end{document}